\documentclass[preprint,3p,authoryear,a4paper]{elsarticle}
\usepackage{graphicx}
\usepackage{epstopdf}
\usepackage{color}
\usepackage{dsfont}
\usepackage{amsmath}
\usepackage{amssymb}
\usepackage{amsfonts}
\usepackage{amsbsy}
\usepackage{caption}
\usepackage{subcaption}
\usepackage{amsthm}
\usepackage{array}
\usepackage{booktabs} 
\usepackage{verbatim}
\usepackage[english]{babel}
\usepackage{natbib}
\usepackage{ulem}
\usepackage{url}
\usepackage{multirow}
\usepackage{subcaption}
\usepackage{float}
\usepackage{numprint}

\renewcommand\appendix{\par
\setcounter{section}{0}
\setcounter{subsection}{0}
\setcounter{table}{0}
\setcounter{figure}{0}
\gdef\thetable{\Alph{table}}
\gdef\thefigure{\Alph{figure}}
\gdef\thesection{\Alph{section}}
\setcounter{section}{0}}

\newcommand{\E}[1]{\ensuremath{\mathbb{E}\left[#1\right]}}
\newcommand{\coml}[1]{\ensuremath{\tilde{\lambda}_{#1}^\parallel}}
\newcommand{\unil}[1]{\ensuremath{\tilde{\lambda}_{#1}^\perp}}
\newcommand{\comt}[1]{\ensuremath{\tau_{#1}^\parallel}}
\newcommand{\unit}[1]{\ensuremath{\tau_{#1}^\perp}}
\newcommand{\comx}[1]{\ensuremath{X_{#1}^\parallel}}
\newcommand{\unix}[1]{\ensuremath{X_{#1}^\perp}}
\newcommand{\comr}[1]{\ensuremath{\rho_{#1}^\parallel}}
\newcommand{\unir}[1]{\ensuremath{\rho_{#1}^\perp}}
\newcommand{\comj}[1]{\ensuremath{J_{#1}^\parallel}}
\newcommand{\unij}[1]{\ensuremath{J_{#1}^\perp}}
\newcommand{\comtheta}[2]{\boldsymbol{\theta_{#1}^\parallel(#2)}}

\newcommand{\cond}[0]{\ensuremath{\left|\right.}}
\newcommand{\cov}[1]{\ensuremath{\text{Cov}\left(#1\right)}}
\newcommand{\var}[1]{\ensuremath{\text{Var}\left(#1\right)}}
\newcommand{\bo}[1]{\ensuremath{\boldsymbol{#1}}}

\newtheorem{theorem}{Theorem}[subsection]
\newtheorem{example}{Example}

\newtheorem{definition}[theorem]{Definition}
\newtheorem{assumption}[theorem]{Model assumption}
\newtheorem{remark}{Remark}[section]
\numberwithin{equation}{section}

\newenvironment{arcitem}{
\begin{list}{---}{
\topsep=1pt
\itemsep=1pt 
\parsep=0pt 
\leftmargin=19pt 
}}
{\end{list}}

\newcounter{arclist}

\newcounter{arcenum}

\newcommand{\rev}[1]{\textcolor{black}{#1}}
\newcommand{\revv}[1]{\textcolor{black}{#1}}

\begin{document}

\normalem

\begin{frontmatter}

\title{\revv{On the modelling of multivariate counts with Cox processes and dependent shot noise intensities}}

\author[UMelb]{Benjamin Avanzi}
\ead{b.avanzi@unimelb.edu.au}

\author[UNSW]{Greg Taylor}
\ead{gregory.taylor@unsw.edu.au}

\author[UNSW]{Bernard Wong}
\ead{bernard.wong@unsw.edu.au}

\author[UNSW]{Xinda Yang\corref{cor}}
\ead{xinda.yang@unsw.edu.au}

\cortext[cor]{Corresponding author. }

\address[UMelb]{Centre for Actuarial Studies, Department of Economics, University of Melbourne VIC 3010, Australia}
\address[UNSW]{School of Risk and Actuarial Studies, UNSW Australia Business School, UNSW Sydney NSW 2052, Australia}

\begin{abstract}
\revv{In this paper, we develop a method to model and estimate several, \emph{dependent} count processes, using granular data. Specifically, we develop a multivariate Cox process with shot noise intensities to jointly model the arrival process of counts (e.g. insurance claims). The dependency structure is introduced via multivariate shot noise \emph{intensity} processes which are connected with the help of L\'evy copulas. In aggregate, our approach allows for (i) over-dispersion and auto-correlation within each line of business; (ii) realistic features involving time-varying, known covariates; and (iii) parsimonious dependence between processes without requiring simultaneous primary (e.g. accidents) events.}




\revv{The explicit incorporation of time-varying, known covariates can accommodate characteristics of real data and hence facilitate implementation in practice. In an insurance context, these could be changes in policy volumes over time, as well as seasonality patterns and trends, which may explain some of the relationship (dependence) between multiple claims processes, or at least help tease out those relationships.}

\revv{Finally, we develop a filtering algorithm based on the reversible-jump Markov Chain Monte Carlo (RJMCMC) method to estimate the latent stochastic intensities and illustrate  model calibration using real data from the AUSI data set.}
\end{abstract}

\end{frontmatter}
{\centering \large}

\section{Introduction}\label{sec:intro}

\subsection{Background}

This paper is concerned with \revv{the multivariate modelling of counts, which are important in many actuarial applications. For instance,} 
\revv{the distribution of aggregate losses would not only depend on the marginal loss distributions, but also on their dependency structure. Dependencies can manifest themselves in the severity or frequency of claims. By nature, this paper focuses on the latter, since it considers counts only.}



\revv{The actuarial literature on counts processes is vast. Beyond the landmark work of \citet{Lun09,Cra30,Cra55,Buh70}, early extensions of the classical  Poisson process approach include \citet*{Arj89,Nor93,Nor99}, who developped a marked point process approach. There, the arrival process of claims follows a continuous time point process and the claim development is captured by a mark. This framework was implemented in \citet*{AnPl14} and \citet*{Lar07}. \citet*{Jew89,ZhZh09,ZhZh10} also studied the use of Poisson process to model claim arrival. \citet*{AvWoYa16} further extended the study to a marked Cox process framework, where a stochastic intensity introduces over-dispersion and an estimation algorithm is developed with the presence of reporting delays. \citet*{BaLiTa16a} and \citet*{BaLiTa19} have also adopted a Cox process approach where the intensity  follows  a hidden Markov model (HMM) with Erlang state-dependent distributions. \citet*{AvTaWoXi20b} model and analyse claim counts of the AUSI dataset using a univariate Markov modulated non homogeneous Poisson process. Discrete time frameworks were also considered \citep*[see, for example][where numbers of claims follow Poisson distributions]{PiAnDe13,PiAnDe14}. Furthermore, \citet*{MaMi10} establish the framework of using a Poisson cluster model in modelling insurance claim process, which is extended to the case of multiple clusters by \citet*{Ma15} and to non-homogeneous observations by \citet*{Ma14}. \citet*{JeMiSa11} adopt a Poisson cluster model for the payment counts and the total loss amounts of claims with real data illustration.} 

\subsection{Motivation and contributions}

\revv{The availability of granular data (also called ``micro-level'' data) is an important motivation for the developments in this paper. When aggregate (``macro-level'') data is used, the small(er) size of the data set may lead to high parameter uncertainty and hence reduce the predictive power of the model. In particular, one can reasonably question whether there is enough data to fit a dependency structure (when required) to a reasonable level of confidence. An aggregate dataset by nature may eliminate some information about useful factors (such as, in a reserving context, seasonality factors in annual triangles). Besides, aggregation does not allow for a separate treatment of different components of the claims processes (such as information on the claim arrival process and individual claim development process). Those components may have different effects on the quantity of interest, and may also change over time in different ways. In short, the forces that drive the quantities of interest are hidden behind a black box in a macro-level model, which is a problem when the environment changes.}

\revv{Two recurrent features that can be commonly observed in insurance data are overdispersion and autocorrelation within loss processes; see, for example, \citet{DeHe08} and \citet*{DeMaPiWa07}. In fact, one could argue that this is likely to arise from frequency rather than severity. This motivates our focus, in this paper, on count processes. The traditional Poisson process approach cannot introduce dependence within marginal processes without a stochastic intensity. Such a process is Cox by definition. This is developed extensively in \citet*[in a univariate framework]{AvWoYa16}.}

\revv{The Cox approach has further strengths in modelling dependence between processes. In the (multivariate) Poisson approach, dependence between processes can only occur due to simultaneous primary events (e.g. accidents involving 2 cars, or a hailstorm affecting multiple vehicles). While this might be appropriate in very rare circumstances, in general such an assumption is too unrealistic and restrictive, and hence will be unable to capture dependence not involving directly observable primary events. The Cox approach relaxes this restriction by modelling dependence via  stochastic \emph{intensity} processes instead. In this paper, we further take advantage of the L\'evy copula approach to introduce dependence across processes through that stochastic intensity. In so doing, we achieve dependence \emph{within} and \emph{between} processes in a parsimonious way (a L\'evy copula might involve just one parameter). The challenge, however, is to fit such a model to a dataset, as the intensity is not observed (only counts). In this paper, we develop a complete methodology for fitting by developing an EM algorithm with a reversible jump Markov (``RJMCMC'') filter.} 

\revv{We illustrate our approach with real data. To better match the characteristics of real data, our approach also explicitly incorporates time-varying, known covariates. In an insurance context, these could be changes in policy volumes over time, as well as seasonality patterns and trends, which may explain some of the relationship between multiple claims processes (or at least help tease out those relationships). Such allowances are vital to enable practical implementation, as such covariates are readily observable in real data.}

\revv{\begin{remark}A prominent example of ``micro-level'' approaches lies in the area of reserving, which may advantageously use the information of individual claims \citep*{AnPl14,AvWoYa16}. Typical micro-level datasets include information such as inception and expiry dates of each policy, the arrival and reporting dates of each claim and the history of transaction and settlement (if the claim has been settled) of each claim. Such data sets are very granular and complex. 
They allow the observations of varying components, such as the arrival and reporting dates as well as the transaction records and settlement status of each claim. Therefore, one can try to build a dependency model with micro-level components to explain the dependencies of the reserving estimate. We will build on this example in the illustration part of this paper. Note, however, that while the reserving application is used as a prominent example in this paper, a complete micro-level reserving model framework would include not only projection of ultimate claim frequencies, but also reporting delay distributions, claim severities with claim development patterns. The methodology of introducing and estimating the reporting delays has been studied in \citet*{AvWoYa16}. Under the assumption of independence between cross Line of Business (``LoB") reporting delays (which we believe to be a good assumption for attritional claims as modelled in this paper), the introduction of reporting delay patterns in the multivariate context is essentially the same as that in the univariate context, where an independent reporting delay distribution is chosen for each process. Thus the estimation would follow the same procedures as that in \cite*{AvWoYa16}.\end{remark}}


\subsection{Structure of the paper}

\revv{In Section \ref{sec:model} we set out our multivariate counts process. Dependence is introduced with the help of a L\'evy copula, which is applied on the shot noise intensity processes of the marginal counts processes. The estimation (and prediction) of the model parameters is addressed in Section \ref{sec:estimationandprediction}, where we develop an EM algorithm with a reversible jump Markov (``RJMCMC'') filter. This is where the availability of granular data is critically useful. Section \ref{S_illustration} presents a comprehensive illustration using a bivariate motor insurance sub-set of the the AUSI data set. We show how the data is detrended by accounting for exposure and further covariate adjustments. The multivariate claim arrival process is built in a bottom up fashion by ``assembling'' two marginal claims arrival processes with the help of a Clayton L\'evy copula. The final model is used to perform predictions and results are thoroughly discussed. Section \ref{S_conclusion} concludes.
}

\section{A multivariate claim arrival process}\label{sec:model}

In this section, we develop our multivariate count model. Section \ref{sec:multiarrival} introduces a multivariate Cox process model with common shock intensities. Section \ref{sec:levycopula} provides an alternative bottom-up construction of a common shock model with a L\'evy copula approach. 

\subsection{A Cox process model with common shocks}\label{sec:multiarrival}

Let $\bo{N}=\left\{\left(N_1(t),\ldots,N_G(t)\right),t\geq 0\right\}$ denote the multivariate claim arrival process. The $g^{th}$ marginal process, $\{N_g(t), t\geq 0\}$, is a counting process for the number of claims of the $g^{th}$ ($1\leq g\leq G$) \revv{counts process}. We use the definition of the multivariate Cox process from \cite*{MoWa04}, and assume that $\bo{N}$ is a multivariate Cox process driven by a multivariate intensity $\bo{\tilde{\lambda}}=\left\{\left(\tilde{\lambda}_1(t),\ldots,\tilde{\lambda}_G(t)\right),t\geq 0\right\}$ (see Definition \ref{model:multicox}).

\begin{definition}[A multivariate Cox process]\label{model:multicox}\citep*[see, for example,][]{MoWa04}
	Suppose that $\tilde{\lambda}_g=\{\tilde{\lambda}_g(\xi):\xi\in S\}$, $g=1,\ldots, G$, are non-negative random fields so that for $g=1,\ldots, G$, $\xi\rightarrow \tilde{\lambda}_g(\xi)$ is a locally integrable function (that is, $\tilde{\lambda}_g(\xi)$ is integrable over the space of events) with probability one (where $\xi$ and $S$ refer to the event and the space of events). Conditional on $\bo{\tilde{\lambda}}=\left\{\left(\tilde{\lambda}_1(t),\ldots,\tilde{\lambda}_G(t)\right),t\geq 0\right\}$, suppose that $\{N_g(t), t\geq 0\}_{g=1,\ldots,G}$ are independent Poisson processes with intensity functions $\{\tilde{\lambda}(t), t\geq 0\}_{g=1,\ldots,G}$ respectively. Then $\bo{N}=\left\{\left(N_1(t),\ldots,N_G(t)\right),t\geq 0\right\}$ is said to be a multivariate Cox process driven by $\bo{\tilde{\lambda}}$.
\end{definition}

The dependency structure of a multivariate Cox process results from the multivariate intensity process. Here the multivariate intensity is influenced by the joint movement of the underlying risk regimes of multiple \revv{counts processes}. Furthermore, 
\rev{the conditional independence of the marginal processes assumption provides} a simple mathematical representation and is able to account for most of the dependency across general insurance claims. 

As in \citet*{AvWoYa16} (in a univariate framework), we model the marginal intensity processes of $\bo{\lambda}$ as shot noise processes. This allows us to introduce a dependency structure via common shocks (common shots) across multiple stochastic intensities.

\begin{definition}[A multivariate shot noise process]\label{definition:commonshock}\citep*{Li02} A multivariate stochastic process, $\bo{\tilde{\lambda}}=\left\{\left(\tilde{\lambda}_1(t),\ldots,\tilde{\lambda}_G(t)\right),t\geq 0\right\}$, is a multivariate shot noise process if 
	\begin{equation}
		\tilde{\lambda}_g(t)=\tilde{\lambda}_g(0)e^{-\kappa_gt}+\sum_{j=1}^{J(t)} X_{g,j} e^{-\kappa_g(t-\tau_j)}, \quad t\geq 0,
	\end{equation}
	where $\left\{J(t),t\geq 0\right\}$ is a homogeneous Poisson process of intensity $\rho$ (with $\rho>0$); $\tau_j$ represents the arrival time of the $j^{th}$ \rev{event} of $\left\{J(t),t\geq 0\right\}$, which triggers a jump over the shot noise process $\left\{\tilde{\lambda}_g(t),t\geq 0 \right\}$. The size of the jump is denoted by $X_{g,j}$ (where the subscripts refer to the $g^{th}$ marginal shot noise process and $j^{th}$ shot), which follows an exponential distribution with an expected value of $1/\eta_g$ \rev{if $X_{g,j}>0$}. We assume that the sequence of the $G$-dimensional random variable \rev{$\{\bo{X}_j=(X_{1,j},\ldots,X_{G,j}) \}$ are  independent and identically distributed over $j=1,\ldots,J(t),t\geq 0$}. Furthermore, $X_{g_1,j}$ and $X_{g_2,j}$ $(g_1,g_2=1,\ldots,G)$ can be dependent \rev{for all $j=0,\ldots, J(t)$}. Denote by $f_{\bo{X}}$ the density function of $\bo{X}_j$. The speed of decay is measured by the parameter $\kappa_g$. Note $f_{\bo{X}}$ can consist of a mixture of continuous and discrete components.
\end{definition} 

Definition \ref{definition:commonshock} introduces a common shock dependency structure across multiple shot noise processes. It is worth mentioning that one or more marginals of \rev{$\bo{X_j}$ ($j=1,\ldots, J(t)$)} can be 0. In light of this, one can decompose the arrival of shots of a marginal shot noise process into different subsets, where each subset corresponds to a unique combination of non-zero shots on this marginal stochastic intensity process. In a bivariate context, for example, there can be up to three subsets, where shots can affect either exclusively on one marginal intensity process or simultaneously on both marginal intensity processes. We have provided an example (see Example \ref{exe:bivariate} below) to illustrate how such a decomposition can be achieved and presented the covariance between the marginal shot noise intensities at time $t$ ($t>0$). 

\begin{remark} Here $J(t)$ triggers a multivariate jump, and it is possible to have only one marginal being non-zero (in which case it is a unique jump) or more than one marginal being non-zero (hence a common jump). Hence we write $J$ and not of $J_g$.  This is also why $f_{\boldsymbol{X}}$ does not depend on $g$, since it is a \rev{$G$-dimensional} random variable. Furthermore, $X_{g_1,j}$ and $X_{g_2,j}$ can be dependent for all $g_1$ and $g_2$.
\end{remark}

\begin{remark}
	Interested readers can also refer to \citet*{SeSc18} for an alternative way of defining a multivariate Cox process. In \citet*{SeSc18}, a common stochastic clock is shared by all marginal processes. This requires a different interpretation of the model as well as different estimation techniques.
\end{remark}

\begin{example}[A bivariate example] \label{exe:bivariate} A bivariate shot noise process,
	\begin{equation}
		\bo{\tilde{\lambda}(t)}=(\tilde{\lambda}_1(t),\tilde{\lambda}_2(t)),
	\end{equation}
	can be expressed as
	\begin{equation}
		\begin{aligned}
			\tilde{\lambda}_1(t)&=\tilde{\lambda}_1(0)e^{-\kappa_1t}+\sum_{j=1}^{\unij{1}(t)}\unix{1,j} e^{-\kappa_1(t-\unit{1,j})}+\sum_{j=1}^{\comj{12}(t)}\comx{1:12,j} e^{-\kappa_1(t-\comt{12,j})}\\
			\tilde{\lambda}_2(t)&=\tilde{\lambda}_2(0)e^{-\kappa_2t}+\sum_{j=1}^{\unij{2}(t)}\unix{2,j} e^{-\kappa_2(t-\unit{2,j})}+\sum_{j=1}^{\comj{12}(t)}\comx{2:12,j} e^{-\kappa_2(t-\comt{12,j})}\\
		\end{aligned}
	\end{equation}
	where $\unij{g}(t)$ ($g=1,\ldots, G$) is a homogeneous Poisson process (of intensity $\unir{g}$) that triggers shots of size $\unix{g,j}$ (with distribution function $F^\perp_g$) and arrival time $\unit{g,j}$, only on the $g^{th}$ marginal process (for $g=1,2$); and $\comj{12}(t)$ is a homogeneous Poisson process (of intensity $\comr{12}$) that affects both marginal processes simultaneously with a bivariate shot of size \rev{$(\comx{1:12},\comx{2:12})$} (with a bivariate distribution function $F^\parallel_{12}$ of marginals $F^\parallel_{1:12}$ and $F^\parallel_{2:12}$) and arrival time $\comt{12,j}$, for $j=1,\ldots,\comj{12}(t)$. \rev{Note that we have omitted the unique shots of $\tilde{\lambda}_2(t)$ in the expression of $\tilde{\lambda}_1(t)$ for the ease of notation and also the benefit that $F_1$ is not necessarily a mixed distribution. Equivalently, one can adopt the convention in Definition \ref{definition:commonshock}, in which case some shots can be 0 for a marginal process and hence the distribution consists of a mass at 0.}

	\begin{figure}[htb]
		\centering
		\includegraphics[width=.6\textwidth]{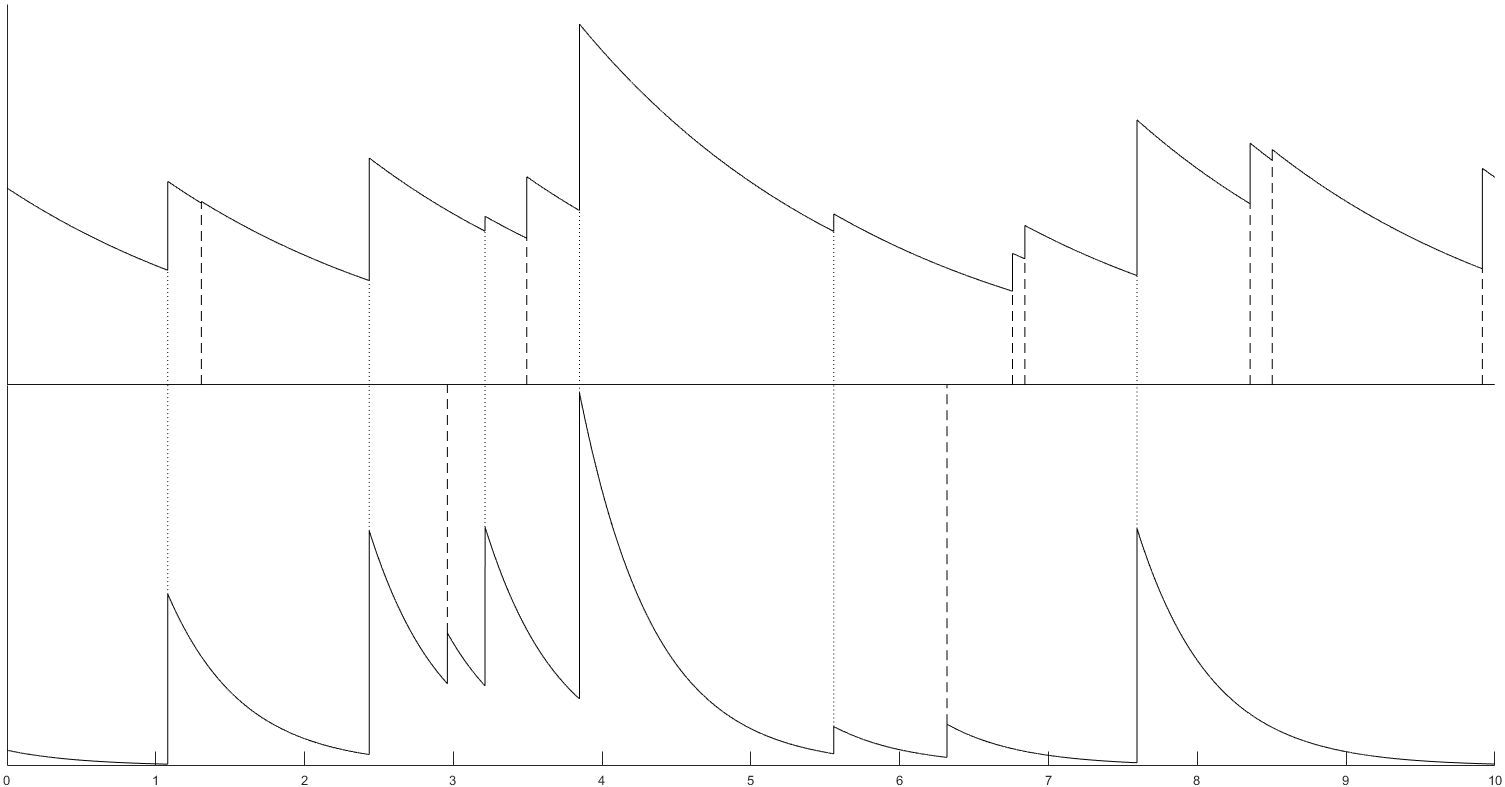}
		\caption{An illustration of a bivariate shot noise process: dotted lines - common shots, dashed lines - unique shots.}
		\label{fig:shot_multi}
	\end{figure}
	
	Figure \ref{fig:shot_multi} illustrates a realisation of the path of a bivariate shot noise process. The dotted lines indicate common shots which trigger jumps in both marginal shot noise processes. Moreover, the dashed lines indicate unique shots, where each shot only triggers jumps in one marginal process.
	
	Furthermore, 
	\begin{equation}\label{eq:cov_shotnoise}
		\cov{\tilde{\lambda}_1(t),\tilde{\lambda}_2(t)}=\frac{\comr{12}\E{\comx{1:12}\comx{2:12}}}{\kappa_1+\kappa_2}
	\end{equation}
	where $(\comx{1:12},\comx{2:12})$ follows a bivariate distribution function $F^\parallel_{12}$\rev{; see Appendix \ref{append:proof} for a proof.} Finally, the marginal process $\tilde{\lambda}_1(t)$ is a univariate shot noise process, that is,
	\begin{equation}
		\tilde{\lambda}_1(t)=\tilde{\lambda}_1(0)e^{-\kappa_1t}+\sum_{j=1}^{J_1}X_{1,j} e^{-\kappa_1(t-\tau_{1,j})},
	\end{equation}
	where $J_1$ is a homogeneous Poisson process of intensity 
	\begin{equation}
		\rho_1=\unir{1}+\comr{1},
	\end{equation}
	and \rev{where} the size of shot $X_{1,j}$ follows the distribution function
	\begin{equation}
		F_1=\frac{\unir{1}}{\rho_1}F^\perp_1+\frac{\comr{1}}{\rho_1}F^\parallel_{1:12}.
		\label{eq:pdf_size}
	\end{equation}
\end{example}

One can further generalise Example \ref{exe:bivariate} to \rev{higher dimensions} in a similar manner. In general, there can be up to $2^G-1$ subsets for a $G$-dimension shot noise process. 

The frequencies of insurance claims are also driven by risk exposures. The exposure an insurance process faces depends on \rev{factors such as the volumes of business, and can sometimes} be measured by the number of policies in force, for instance. Furthermore, there are also trends and seasonal patterns that affect the frequencies of claims. In a tropical region, for example, one would expect a higher number of claims in summer for a housing insurance product than winter. We incorporate risk exposure into the multivariate Cox process $\bo{N}$ through $\{W_g(t)\}_{g=1,\ldots,G}$, where $W_g(t)$ is the risk exposure of the $g^{th}$ \revv{count process} at time $t$\rev{; see also Section \ref{sec:fitting_covariates}}.

\begin{assumption}[A non-stationary multivariate shot noise intensity]\label{model:multishotnoise}
	We assume that there exists a stationary multivariate shot noise process $\bo{\tilde{\lambda}(t)}$ such that for $g=1,\ldots,G$
	\begin{equation}
		\lambda_g(t)=W_g(t)\tilde{\lambda}_g(t).
	\end{equation}
\end{assumption}

Assumption \ref{model:multishotnoise} allows for a multivariate intensity process \revv{for the arrivals}. The common shock structure can be interpreted as common adverse events that affect multiple \revv{counts processes} at different scales. Furthermore, we assume that the stochastic intensity of \revv{arrivals} is proportional to the risk exposure of the corresponding \revv{counts process}, which leads to a non-stationary shot noise process. Here a stationary shot noise process is a process that follows Definition \ref{definition:commonshock}, where the joint distribution of the process does not depend on $t$, and the only source of non-stationarity is introduced through the risk exposure.

\begin{remark}
    	\rev{As the exposure process is non-stationary and the underlying intensity process is unobservable, the analysis of empirical moments (including autocorrelations) may be distorted. As such, direct evaluation of whether a Cox model is a good candidate is difficult. Instead, in this paper the choice of candidate is driven by theoretical properties and goodness-of-fit analysis of the model (as illustrated later in this paper with residual analysis). In our illustration, we investigated and were able to explain the non-stationary patterns of the claim intensity process resulting from exposure, seasonal patterns and trends.}
\end{remark}

\subsection{A bottom-up approach of common shot constructions with L\'evy copulas}\label{sec:levycopula}

Definition \ref{definition:commonshock} defines a general common shock model. This allows for dependency construction of higher dimensions, which is an extension of Example \ref{exe:bivariate}. However, \rev{such a} common shock representation of the dependency structure faces several challenges. 
Firstly, the practical application \revv{of counts processes in actuarial contexts (more generally, compound loss processes)} typically adopts a bottom-up approach  in model construction and calibration. Such an approach involves the modelling and estimation of the \revv{marginal processes} before a dependency structure is introduced \citep*[see page 221,][for the definition of a bottom-up approach]{McFrEm15}. However, this is not directly available to a common shock model, where the specification of the dependency structure is \emph{implied} in a multivariate set-up.
Secondly, the direct construction of a common shock model may lead to complexity of model assumptions and a high number of parameters. This is particularly relevant at a high dimension where there will be a number of $2^G-1$ arrival processes of shots. Furthermore, each common shock process requires a dependency model of the joint jump sizes, which effectively requires a number of $2^G-G-1$ dependency structures \rev{\citep[further discussion of the use of common shock models can be found in, e.g.,][]{LiMc03}. This is clearly not practical, and because the joint intensity process is unobservable, over-parametrisation may further lead to higher parameter errors and model errors.} 

\rev{Fortunately, one can achieve an identical outcome in parsimonious way thanks to L\'evy copulas; refer to \citet*{Tank03} and \citet*[in the actuarial literature]{AvCaWo11} for the formal definition and discussion for the properties of L\'evy copulas. A L\'evy copula model typically includes only a small number of parameters (sometimes just one) and hence leads to a generally parsimonious model specification. Furthermore,} a L\'evy copula approach separates the model specification of the marginal processes and the dependency structure, which allows for a bottom-up approach in model construction. In Section \ref{sec:estimationandprediction}, we will further explain how one can separate the calibration of the marginal components and the dependency component.  \rev{For all those reasons} we adopt the L\'evy copula approach, which provides a \rev{practical and} parsimonious approximation to the underlying joint intensities.

\rev{Under Definition \ref{definition:commonshock}, a large range of L\'evy copulas can specify the dependency structure across the marked Poisson process (that is, arrival of shot \emph{and} the associated jump sizes). Such a dependency relationship drives the dependency across the marginal Cox processes. A real world interpretation of this mechanism is that dependency of multiple insurance count processes do not arrive directly from common arrival of claims - rather, it is the dependency in the underlying risk generating regime (e.g. the intensity processes in this paper) that creates the dependency of claim counts. Continuing on Example \ref{exe:bivariate}, Example \ref{exo-clayton} introduces the bivariate Clayton L\'evy copula, which we will use in Section \ref{S_illustration}.}

\begin{example}
\label{exo-clayton}
\rev{A bivariate L\'evy copula $\mathfrak{C}$ can be used to couple the marginal processes together via
\begin{equation}
	U(x_1,x_2)=\mathfrak{C}(U_1(x_1),U_2(x_2)),
\end{equation}
where $U_g(x_g)=\rho_g(1-F_g(x_g))$ denotes the $g^{th}$ tail integral of the marginal process $(g=1,2)$ and $U(x_1,x_2)=\rho_{12}(1+F_{12}^\parallel(x_1,x_2)-F_{12}^\parallel(x_1,\infty)-F_{12}^\parallel(\infty,x_2))$ denotes the bivariate tail integral of the joint bivariate process.}

\rev{A possible choice for $\mathfrak{C}$ is the bivariate Clayton L\'evy copula \citep*{CoTa04}, which is defined as 
	\begin{equation}
		\mathfrak{C}(u_1,u_2)=(u_1^{-\delta}+u_2^{-\delta})^{-1/\delta},\;\delta>0.
	\end{equation}
It is worth emphasizing that, while there is a relationship between the copula of the common jumps induced by a Clayton L\'evy copula, and the Clayton distributional copula, a Clayton L\'evy copula is \emph{not} a Clayton copula. A Clayton copula is a copula function for a multivariate random variables and a Clayton L\'evy copula (which is not a copula function) aggregates multiple L\'evy processes.}
\end{example}

\section{Parameter estimation and prediction}\label{sec:estimationandprediction}

In this section, we explain how one can estimate the parameters of the multivariate shot noise model. This is particularly essential with a Cox process approach, where the joint intensity is unobservable and hence the model cannot be calibrated with a maximum likelihood approach. Furthermore, the estimation of the parameters and the underlying intensity are necessary for the projection of future claims counts \revv{(e.g., to help with the valuation of future claims liabilities). For the rest of this section, we use a trivariate Cox process as an example.}

We start by introducing notation and likelihood in Section \ref{S_NotLik}, before developing an EM algorithm with a reversible jump Markov chain Monte Carlo (``RJMCMC") filter. We extend the estimation algorithm in \citet*{AvWoYa16} to allow for a multivariate shot noise process. We explain our filtering algorithm with fixed parameters in Section \ref{sec:rjmcmc} and develop a three-step EM algorithm to update the parameter estimates in Section \ref{sec:multiem}. For simplicity, we illustrate the filtering and parameter estimation procedures in a trivariate context. These can be adapted in higher dimensions with a similar procedure.

\subsection{Notation and likelihood} \label{S_NotLik}
Suppose that the overall observation period (that is, the period for which which policy and claim data are extracted) is $[0,T]$, which is discretised into a number of $L$ sub-periods of equal length $\Delta$. We say that a claim arrives in the $i^{th}$ accident period if the arrival time of the claim falls into $((i-1)\Delta,i\Delta]$. This set of observations is denoted 
\begin{equation}
	\mathbf{N_D^G}=\{N_{g,i}; g=1,2,3,\; i=1,2,\ldots, L\}.
\end{equation}
Furthermore, we characterise the trajectory of the trivariate shot noise process by 
\begin{equation}
	\boldsymbol{\theta_G}=\{(\tilde{\lambda}_1(0),\tilde{\lambda}_2(0),\tilde{\lambda}_3(0)),\tau_1,\ldots,\tau_n,\mathbf{X}_1,\ldots,\mathbf{X}_n\}
\end{equation}
where $\tau_i$ and $\mathbf{X}_i$ denote the arrival time and trivariate sizes of the $i^{th}$ shot, respectively. As mentioned in the explanation for Definition \ref{definition:commonshock}, each $\bo{X}_i$ is a trivariate random variable where one or more marginal can be 0. Note that we are modelling the \emph{intensity} of the claim count process here (not the counts themselves), so that all components of $\boldsymbol{\theta_G}$ are unobservable.

In the rest of this section, we will derive the log-likelihood functions for the trajectory of a trivariate shot noise process and the conditional observations of claim counts given the shot noise trajectory. 

Denote by $\boldsymbol{\delta}$ the parameter vector of the L\'evy copula. \revv{Furthermore, denote by $\rho_j$ ($j=1,2,3$) the intensity of the Poisson process of the non-zero shots of the $j^{th}$ margin. Note that $\rho_j$ ($j=1,2,3$) is the thinned Poisson process (conditional on positive shots) of the $j^{th}$ margin of the trivariate Poisson arrival of shots with intensity $\rho$ (see Definition \ref{definition:commonshock}).} The log-likelihood of $\boldsymbol{\theta_G}$ is
\begin{equation}\label{eq:llh_prior}
	\begin{aligned}
		& \log p(\boldsymbol{\theta_G};\rho_1,\rho_2,\rho_3,\eta_1,\eta_2,\eta_3,\kappa_1,\kappa_2,\kappa_3,\boldsymbol{\delta})\\
		=& n\log \revv{\rho}-\revv{\rho}T+\log \revv{f_{\mathbf{X}}}(\mathbf{X_n})\\
		&+\sum_{g=1}^{3}\left[\left(\dfrac{\rho_g}{\kappa_g}-1\right)\log\tilde{\lambda}_g(0)-\eta_g\tilde{\lambda}_g(0)+\dfrac{\rho_g}{\kappa_g}\log\eta_g-\log\left(\Gamma\left(\dfrac{\rho_g}{\kappa_g}\right)\right)\right],
	\end{aligned}
\end{equation}
where each term in the summation refers to the log-likelihood of the corresponding initial value of a marginal shot noise process, which is chosen to be the corresponding stationary distribution of each univariate process, that is, a Gamma random variable \citep*[see][]{CeMi06a,CeMi06}, and where $\Gamma(\cdot)$ is the Gamma function. Conditional on the trajectory of the multivariate shot noise process, the conditional log-likelihood of our observations is

\begin{equation}\label{eq:llh_cond}
	\begin{aligned}
		& \log L(\boldsymbol{N_D^G}|\boldsymbol{\theta_G};\kappa_1,\kappa_2,\kappa_3)\\
		=& \sum_{g=1}^{3}\sum_{i=1}^{L}\log A\left(N_{g,i},M_{g,i};\kappa_g\right)\\
		=&\sum_{g=1}^{3}\sum_{i=1}^{L}\left(-M_{g,i}+N_{g,i}\log M_{g,i}\right)+\text{constant}		
	\end{aligned}
\end{equation}
where 

\begin{equation}
	\begin{aligned}
		M_{g,i}&=\int_{(i-1)\Delta}^{i\Delta}\lambda_g(t)\;\mathrm{d}t,\\
		A(N_{g,i},M_{g,i};\kappa_g)&=\dfrac{e^{-M_{g,i}}M_{g,i}^{N_{g,i}}}{N_{g,i}!}
	\end{aligned}
\end{equation}

Here the function $A(m,M_{g,i};\kappa_g)$ calculates the probability of having $m$ ultimate claims, which comes from a Poisson probability mass function of intensity $M_{g,i}$. This function depends on $\kappa_g$ through the integral of $\lambda_g(t)$. Note that the format of the conditional likelihood function \eqref{eq:llh_cond} results from the discretisation scheme introduced at the beginning of this section. Such a discretisation scheme accommodates the discrete nature of real data, which is an essential step in applying a continuous time Cox process \citep*[see also][]{AvWoYa16}.

\subsection{A RJMCMC filtering algorithm}\label{sec:rjmcmc}

The issue we face is to derive the conditional distribution of the unobservable component, $\boldsymbol{\theta_G}$, given the knowledge of $\boldsymbol{N_D^G}$ (which is observable), which is a filtering problem \citep*{CeMi06a,CeMi06}. Since we can obtain the likelihood of the shot noise process itself---see Equation \eqref{eq:llh_prior}---and the likelihood of the conditional distribution of $\boldsymbol{N_D^G}$ given $\boldsymbol{\theta_G}$---see Equation \eqref{eq:llh_cond}, we can adopt a MCMC algorithm. However, since the number of shots is unknown, the dimension of $\boldsymbol{\theta_G}$ is also undetermined. We thus adopt a RJMCMC algorithm \citep*{Gre95} to allow for `moves' with dimension changing. We will briefly outline the general steps of a RJMCMC simulation algorithm. One can refer to \citet*[][Chapter 3]{GeJoBrMe11} for more details about the RJMCMC simulation and \citet*{CeMi06a,CeMi06} and \citet*{AvWoYa16} regarding using a RJMCMC filter in the univariate case of a shot noise Cox process. 

A RJMCMC simulation algorithm helps approximate the conditional distribution of the shot noise process given the observations. Firstly, one randomly chooses a move type with probability $p(r|n)$ (with $\sum_rp(r|n)=1$), which depends on the existing number of shots. Given a chosen move type $r$ and the existing shot noise process $\boldsymbol{\theta_G}$, one proposes a new shot noise trajectory by generating a random component $\mathbf{u}$ with a proposal distribution $q(\mathbf{u}|r,n,\boldsymbol{\theta_G})$. This proposal, once accepted, results in $\boldsymbol{\theta_G'}$ with $n'$ shots. The probability of accepting the proposal is $\min(1,\alpha[(n,\boldsymbol{\theta_G}),(n',\boldsymbol{\theta_G'})])$, where $\alpha[(n,\boldsymbol{\theta_G}),(n',\boldsymbol{\theta_G'})]$ is calculated according to \eqref{eq:acc_ratio} and can be further decomposed into the product of the likelihood ratio, prior ratio, proposal ratio, and Jacobian.

\begin{equation}\label{eq:acc_ratio}
	\alpha[(n,\boldsymbol{\theta_G}),(n',\boldsymbol{\theta_G'})]=\min\left\{1,
	\underbrace{L(\boldsymbol{N_D^G}|\boldsymbol{\theta_G'})\over L(\boldsymbol{N_D^G}|\boldsymbol{\theta_G})}_\text{likelihood ratio}
	\times\underbrace{{p(\boldsymbol{\theta_G'}) \over p(\boldsymbol{\theta_G})}}_\text{prior ratio}
	\times\underbrace{{p(r'|n')\over p(r|n)}
		{q(\boldsymbol{u'}|r',n',\boldsymbol{\theta_G'})\over q(\boldsymbol{u}|r,n,\boldsymbol{\theta_G})}}_\text{proposal ratio}
	\times\underbrace{\lvert \dfrac{\partial f_{r,n}(\boldsymbol{\theta_G},\boldsymbol{u})}{\partial(\boldsymbol{\theta_G},\boldsymbol{u})} \rvert}_\text{Jacobian}
	\right\}.
\end{equation}

We introduce five move types, namely s, p, h, b and d. For simplicity, we choose $p(r|n)=0.2$ for $r=$ s, p, h, b, d and $n>1$. Furthermore, we assume that $p(s|0)=p(b|0)=0.5$, that is, a move can only be either of type s or type b (with equal probabilities) if there is no existing shot.
The algorithm is similar to that in \citet*{CeMi06a,CeMi06} and \citet*{AvWoYa16}, except that we now look at the multivariate case. Furthermore, we adopt the dash symbol ($'$) for all variables related to the proposed shot noise trajectory (e.g. $n'$ refers to the number of shots in the proposed state).

We start by introducing three moves that do not involve dimension changing, namely move s, p and h. In other words, we have $n=n'$.

Move s proposes to change the initial values of the multivariate shot noise process. For each of the $g^{th}$ marginal shot noise process ($g=1,2,3$), the new initial value, $\left(\tilde{\lambda}_g'(0)\right)$, is drawn from the Gamma distribution with parameters $(\rho_g/\kappa_g,\eta_g)$. The prior ratio is calculated as
\begin{equation}
	\prod_{g=1}^{3}e^{-\eta_g\left(\tilde{\lambda}_g'(0)-\tilde{\lambda}_g(0)\right)}\left(\frac{\tilde{\lambda}_g'(0)}{\tilde{\lambda}_g(0)}\right)^{\frac{\rho_g}{\kappa_g}-1}
\end{equation}
and the proposal ratio as
\begin{equation}
	\prod_{g=1}^{3}e^{-\eta_g\left(\tilde{\lambda}_g(0)-\tilde{\lambda}_g'(0)\right)}\left(\frac{\tilde{\lambda}_g(0)}{\tilde{\lambda}_g'(0)}\right)^{\frac{\rho_g}{\kappa_g}-1}.
\end{equation}
Move p proposes to change the position of an existing shot. This involves choosing an existing shot by generating a value, denoted by $n^*$, from the discrete uniform distribution over $\{1,\ldots,n\}$. Then the position of the $n^{*th}$ shot, $\tau_{n^*}$, is proposed to be changed to $\tau'_{n^*}$. The proposal follows a continuous uniform random distribution over $(\tau_{n^*-1},\tau_{n^*+1})$ where $\tau_0=0$ and $\tau_{n+1}=T$. 
The prior ratio of this move is 1, since the location of the shot is irrelevant to the prior likelihood (see Equation \eqref{eq:llh_prior}). Furthermore, one can also show that the proposal ratio is 1, which is due to the use of the uniform distribution in the proposal.

Move h proposes to change the size (height) of an existing shot. Similar to move p, move h requires selecting an existing shot $n^*$ from the discrete uniform distribution over $\{1,\ldots,n\}$. Then the size of the $n^{*th}$ shot, $\mathbf{X_{n^*}}$, is proposed to be changed to $\mathbf{X'_{n^*}}$. The proposal follows the mixed density function \eqref{eq:pdf_size}. The prior ratio is
\begin{equation}
	\frac{f_{\bo{X}}\left(\mathbf{X'_{n^*}}\right)}{f_{\bo{X}}\left(\mathbf{X_{n^*}}\right)}
\end{equation}
and the proposal ratio is
\begin{equation}
	\frac{f_{\bo{X}}\left(\mathbf{X_{n^*}}\right)}{f_{\bo{X}}\left(\mathbf{X'_{n^*}}\right)}.
\end{equation}

One may notice that the product of the proposal ratio and prior ratio is always 1 for moves s, p and h. This is because we choose the proposal distribution to follow the prior knowledge of the shot noise trajectory (given a fixed number of shots). Furthermore, the Jacobian is always 1 for these three moves since the dimension of $\boldsymbol{\theta_G}$ does not change. 

Now we introduce two moves that change the dimension of $\boldsymbol{\theta_G}$. Firstly, we have move b that gives birth to a new shot. This involves drawing a new position $\tau^*$ from a continuous uniform distribution over $[0,T]$. Denote by $n^*$ the integer such that $\tau_{n^*-1}<\tau^*<\tau_{n^*+1}$ with $\tau_0=0$ and $\tau_{n+1}=T$. In this case, we have $n'=n+1$. Furthermore, the size of this shot, $\mathbf{X'_{n^*}}$, is simulated from the (mixed) density function $f_{\bo{X}}$. Furthermore, we have a move type d that delete an existing shot. This includes drawing $n^*$ from a discrete uniform distribution from $\{1,\ldots,n\}$ and the $n^{*th}$ shot is deleted. Then we have $n'=n-1$. 

For move b, the prior ratio is
\begin{equation}
	\rho f_{\bo{X}}(\mathbf{X'_{n^*}}),
\end{equation}
and the proposal ratio of move b is
\begin{equation}
	\frac{p(d|n+1)}{p(b|n)}\frac{(1+n)^{-1}}{T^{-1}f_{\bo{X}}(\mathbf{X'_{n^*}})}.
\end{equation}

For move d, the prior ratio is
\begin{equation}
	\left(\rho f_{\bo{X}}(\mathbf{X_{n^*}})\right)^{-1},
\end{equation}
and the proposal ratio is
\begin{equation}
	\frac{q(b|n-1)}{q(d|n)}\frac{T^{-1}f_{\bo{X}}(\mathbf{X_{n^*}})}{n^{-1}}.
\end{equation}

One can show that the Jacobian for moves b and d is still 1. The likelihood ratio involved in \eqref{eq:acc_ratio} can be obtained from \eqref{eq:llh_cond}. It is worth mentioning that this is based on the observations of $\mathbf{N_D^G}$ and hence involves the (non-stationary) risk exposure (as part of calculating $M_l$ for $l=1,\ldots,L$).

\rev{We have summarised the various move types in Table \ref{tab:moves}.}
\begin{table}[htb]
	\small
	\centering
	\rev{\begin{tabular}{lp{10cm}l}
		\hline
		move type (`$r$') & proposal of the next state                                                                                                                                                                                                      & $p(r|n)$                 \\ \hline
		s                 & modifying the initial value of $\left(\left(\tilde{\lambda}_1'(0),\tilde{\lambda}_2'(0),\tilde{\lambda}_3'(0)\right)\right)$ by drawing $\left(\tilde{\lambda}_g'(0)\right)$ from the stationary distribution of the $g^{th}$ marginal ($g=1,2,3$)             & 0.5 ($n=0$), 0.2 ($n>0$) \\ \hline
		b                 & generating a new shot by drawing a new position $\tau^*$ uniformly from $(0,t]$ and drawing a new jump height $\mathbf{X'_{n^*}}$ from the shot size distribution $f_{\bo{X}}$                                                 & 0.5 ($n=0$), 0.2 ($n>0$) \\ \hline
		h                 & changing the height of a shot by drawing $j$ from the discrete uniform distribution over $\{1,\ldots,n\}$ and drawing $\mathbf{X'_{n}}$ from the shot size distribution $f_X$ to replace $\mathbf{X_{n}}$                         & 0 ($n=0$), 0.2 ($n>0$)   \\ \hline
		p                 & changing the position of a shot by drawing $j$ from the discrete uniform distribution over $\{1,\ldots,n\}$ and drawing a new position, $\tau'_j$, uniformly over $(\tau_{j-1},\tau_{j+1})$ (where $\tau_0=0$ and $\tau_{n+1}=t$) & 0 ($n=0$), 0.2 ($n>0$)   \\ \hline
		d                 & deleting a shot by drawing $j$ from the discrete uniform distribution over $\{1,\ldots,n\}$ and deleting the $j^{th}$ shot                                                                                                      & 0 ($n=0$), 0.2 ($n>0$)   \\ \hline
	\end{tabular}}
	\caption{Types of moves}
	\label{tab:moves}
\end{table}

\subsection{Parameter estimation and prediction}\label{sec:multiem}

An Expectation Maximisation (``EM") algorithm will iteratively update the parameter estimates with the presence of incomplete observations \citep*[see][for more details]{Ryd96}. In particular, a Monte Carlo Expectation Maximisation (``MCEM") algorithm is adopted where the conditional expectation in the E-step is approximated through simulations. \citet*{AvWoYa16} has used an MCEM algorithm to estimate the shot noise parameters in a univariate scenario. In this case, we have further extended the algorithm to the multivariate case. In particular, we follow the idea of the Inference Functions for Margins ("IFM") in the context of copula fitting \citep*[see Chapter 10 of][for more details]{Joe97} and separate the estimation of the parameters of the univariate components and the dependency structure.

Firstly, one starts with calibrating the marginal Cox processes. For each marginal process, we follow the algorithm of \citet*{AvWoYa16} to estimate the shot noise parameters. Secondly, we proceed to estimating the parameters of the L\'evy copula with a MCEM algorithm while the parameters of the marginal processes are fixed. We will explain the full details of this MCEM algorithm in the rest of this section.

The initial estimates of the L\'evy copula parameters $\bo{\gamma}$, denoted by $\bo{\gamma^0}$, are estimated via moment matching based on Equation \eqref{eq:cov_shotnoise} (see Example \ref{exe:bivariate}). Such a moment matching estimation can be applied to a bivariate L\'evy copula and more generally a higher-order nested Archimedean L\'evy copula. With a higher-order nested Archimedean L\'evy copula, one can calculate the covariances between claim frequencies, which can be used to calculate the parameters given the corresponding bivariate marginal L\'evy copulas \citep*[see][]{AvTaWoYa16}.

The initial estimate of the multivariate trajectory of the unobservable shot noise process is obtained based on the estimates of the marginal intensities. The latest estimate of each marginal shot noise is treated as a multivariate shot noise where the other marginals are 0. Therefore one can combine all the marginal estimates and create a multivariate shot noise where each shot is always unique. Furthermore, shots have been merged as long as the arrival times of two shots are less than a threshold (chosen as 0.01 of a day in this project), where the size of the resulting shot is simply the addition of the sizes two individual shots. Such a procedure creates a reasonable initial estimate that utilises the results of the univariate estimation. An illustration of the above procedure in a bivariate case is provided in Example \ref{tab:illustration}.

\begin{example}	\label{exa:illustration}
In Table \ref{tab:illustration} we illustrate how initial estimates of a bivariate trajectory of the unobservable shot noise process can be obtained based on two marginal intensities. Here, the threshold used is 0.5.

\begin{table}[!htb]

	\begin{minipage}{.5\linewidth}
		\centering
	\begin{tabular}{l|l}
		\hline
		arrival time & size of shot \\ \hline
		0            & 1            \\ \hline
		0.7          & 1.5          \\ \hline
		2.8          & 1.2          \\ \hline
	\end{tabular}
	\caption*{Marginal 1}
\label{tab:exe_1}
	\end{minipage}%
	\begin{minipage}{.5\linewidth}
		\centering
	\begin{tabular}{l|l}
		\hline
		arrival time & size of shot \\ \hline
		0            & 2            \\ \hline
		1.5          & 1            \\ \hline
		3.2          & 1.9          \\ \hline
	\end{tabular}
	\caption*{Marginal 2}
\label{tab:exe_2}
	\end{minipage}

\medskip
\medskip
\medskip

\centering
	\begin{minipage}{.5\linewidth}
	\centering
	\begin{tabular}{l|l|l}
		\hline
		 & \multicolumn{2}{|c}{size of shots} \\ \cline{2-3}	
		arrival time & Marginal 1 & Marginal 2 \\ \hline
		0            & 1                         & 2                         \\ \hline
		0.7          & 1.5                       & 0                         \\ \hline
		1.5          & 0                         & 1                         \\ \hline
		3.2          & 1.2                       & 1.9                       \\ \hline
	\end{tabular}
	\caption*{Initial estimate of the bivariate trajectory}
	\label{tab:exe_3}
\end{minipage} 

	\caption{Example \ref{exa:illustration}; illustration of the initial guess of a bivariate shot noise trajectory}
	\label{tab:illustration}
\end{table}

\end{example}

\noindent Once the initial estimates are obtained, one can follow the following procedure:

\begin{arcitem}
	\item in the $k^{th}$ iteration, generating a large number of RJMCMC iterations given $\bo{\gamma}$ (where $\bo{\gamma^0}$ refer to the initial estimates) based on the algorithm developed in Section \ref{sec:rjmcmc};
	\item approximating the conditional expectation 
	
	\begin{equation}
		Q(\bo{\gamma},\bo{\gamma}^k)=\mathbb{E}_{\bo{\gamma^k}}[\log \mathcal{L}(\boldsymbol{N_D^G},\boldsymbol{\theta_G};\bo{\gamma})|\boldsymbol{N_D}]. 
	\end{equation}
	
	as an average of the conditional likelihoods given the RJMCMC simulations;
	\item deriving the parameter estimates $\bo{\gamma^{k+1}}$ such that $\bo{\gamma^{k+1}}$ maximises the conditional expectation $Q(\bo{\gamma},\bo{\gamma}^k)$.
\end{arcitem}

\section{Illustrative case study - a bivariate motor insurance dataset} \label{S_illustration}

\revv{In this section, we illustrate our model and estimation procedure with a real insurance data set. The data set, which is part of the AUSI data set \citep*[see][for the details on the data set]{AvTaWo16}, consists of observations from a Motor insurance portfolio of a major Australian general insurer. We segment the Motor insurance portfolio by states and illustrate how our methodology can be used to understand the dependency of the claim counts between the states of New South Wales (``NSW") and Victoria (``VIC"). 
We choose NSW and VIC because they are two adjacent large states in Australia (in terms of exposure), and contribute to more than half of the national population. Here we use observations of insurance claims and policy information from 01/January/2006 to 31/December/2010. We performed the following manual adjustments to the data set:}
\begin{arcitem}
	\item We excluded claims resulting from catastrophe events. Although catastrophe events explicitly contribute to the dependency of insurance claims in the neighbourhood states, this source of dependency can be observed \rev{and hence explicitly modelled. Indeed}, the modelling of catastrophe events \rev{benefits from} knowledge that is beyond the actuarial field, and is typically conducted with separate catastrophe models in practice \citep*[see e.g.][]{GrKu05}. In the AUSI data set, catastrophe claims are identified with non-empty catastrophe flags;
	\item We excluded policy records of non-positive gross premium, total sum insured or total excess;
	\item We excluded invalid policy records (for example, where inception dates are after expiry dates and/or inception dates are no earlier than year 9999). This only corresponds to a negligible portion of the data set.
\end{arcitem}

In Section \ref{sec:fitting_covariates}, we explain how we allow for covariates in modelling claims counts. Sections \ref{sec:fitting_univariate} and \ref{sec:fitting_bivariate} illustrate the procedures and provide the results of the univariate and bivariate fitting.

\subsection{Exposure and covariates adjustment}\label{sec:fitting_covariates}

\begin{figure}[!htb]
	\centering	
	\begin{subfigure}{\textwidth}
		\centering
		\includegraphics[width=\textwidth]{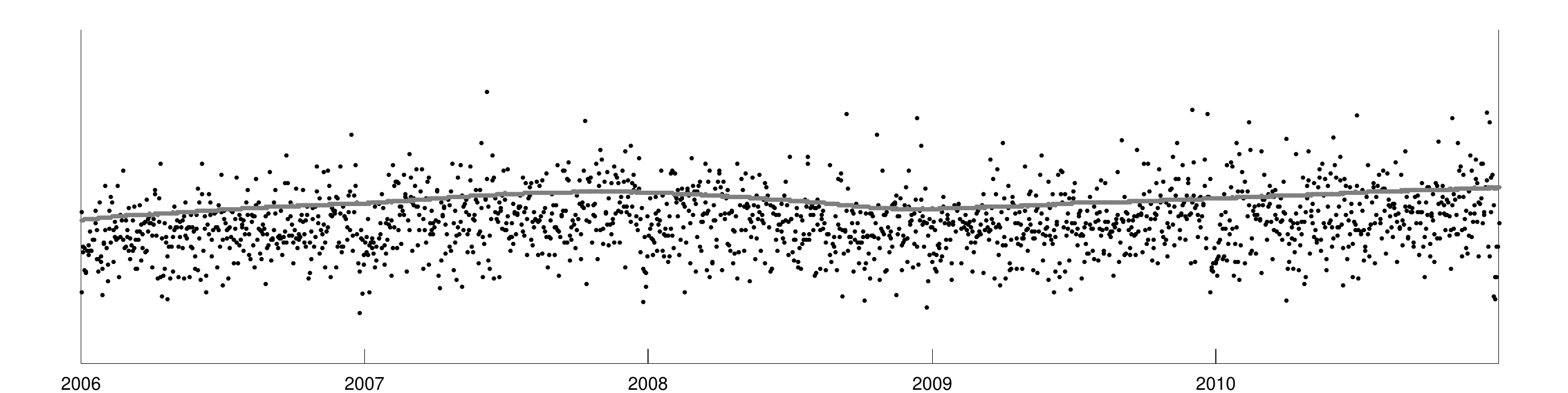}
		\subcaption{Daily claim count (dark) and policy count (grey)}
		\label{fig:nsw_data_ts}
	\end{subfigure}
	
	\begin{subfigure}{0.49\textwidth}
		\centering
		\includegraphics[width=\textwidth]{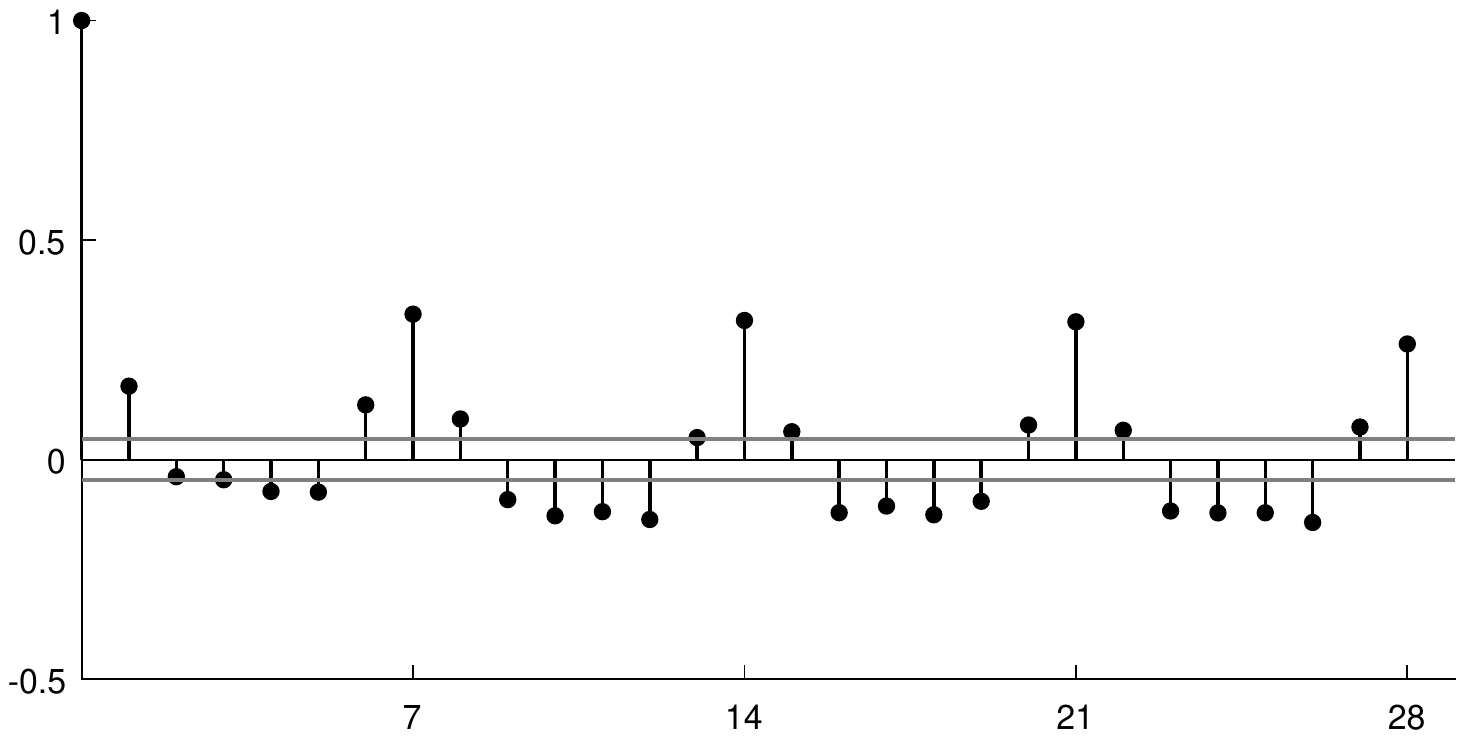}
		\caption{autocorrelation function, daily, raw}
		\label{fig:nsw_data_d} 
	\end{subfigure}
	\begin{subfigure}{0.49\textwidth}
		\centering
		\includegraphics[width=\textwidth]{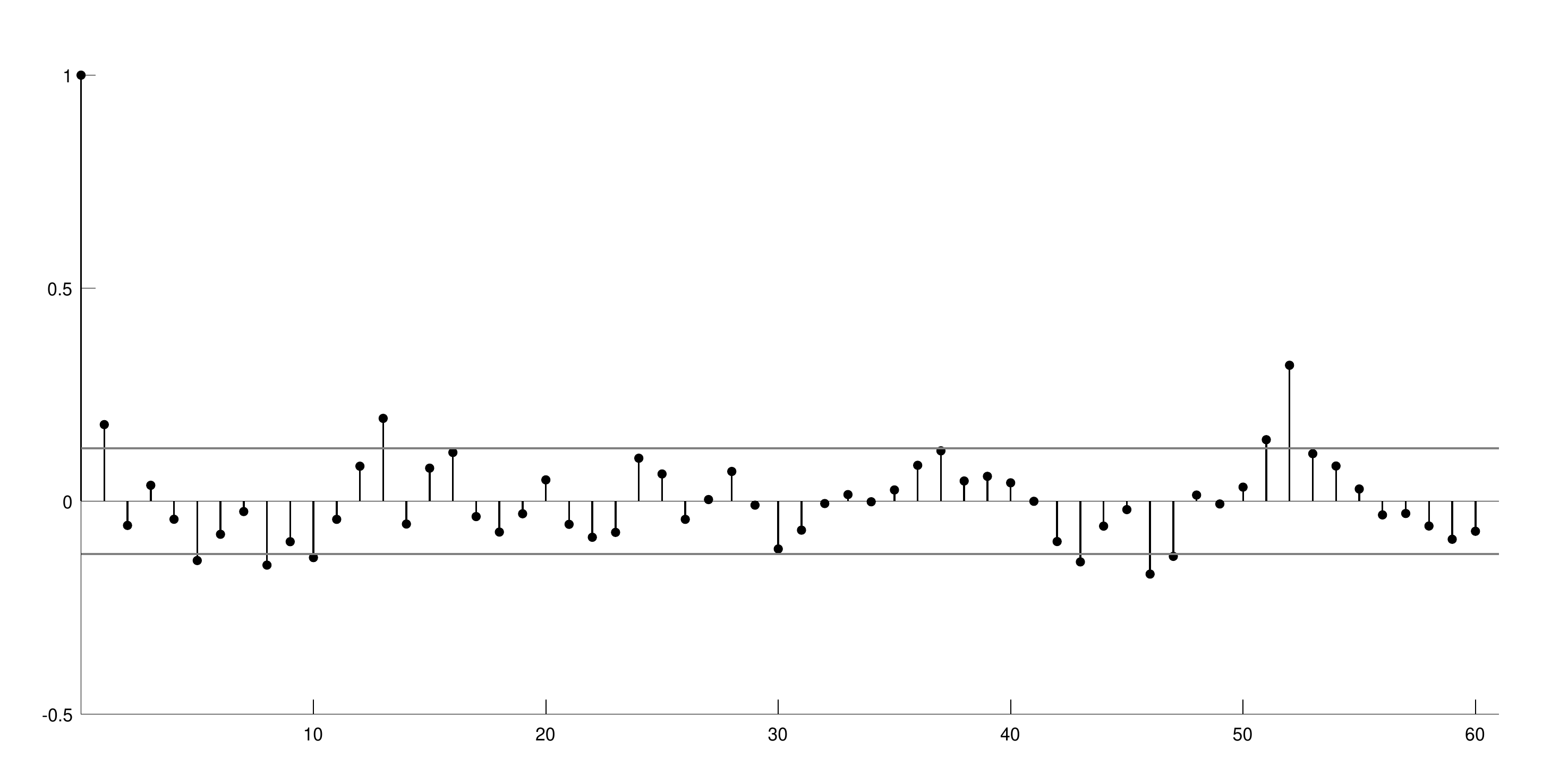}
		\caption{autocorrelation function, weekly, raw}
		\label{fig:nsw_data_w}
	\end{subfigure}

	\begin{subfigure}{0.49\textwidth}
		\centering
		\includegraphics[width=\textwidth]{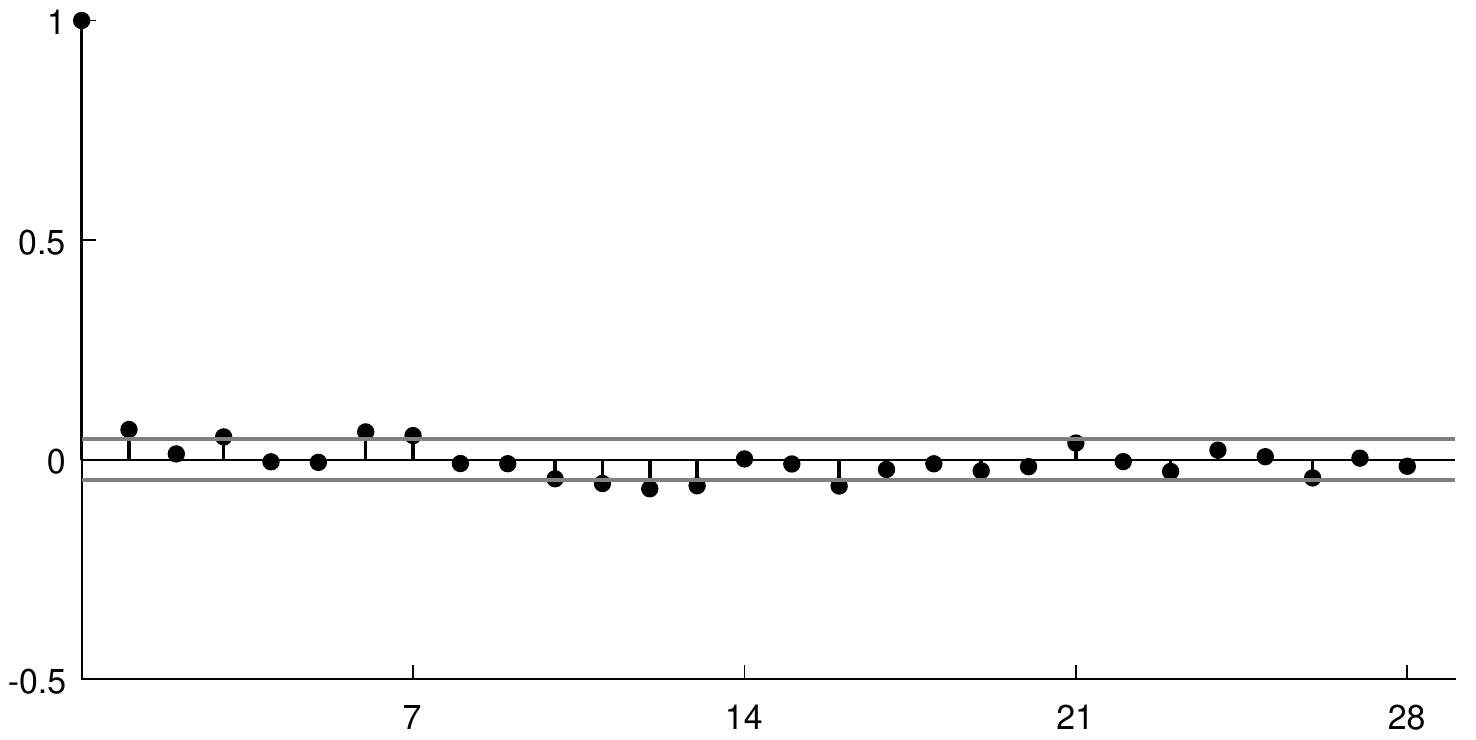}
		\caption{autocorrelation function, daily, after exposure adjustment}
		\label{fig:nsw_data_d_adj}
	\end{subfigure}
	\begin{subfigure}{0.49\textwidth}
		\centering
		\includegraphics[width=\textwidth]{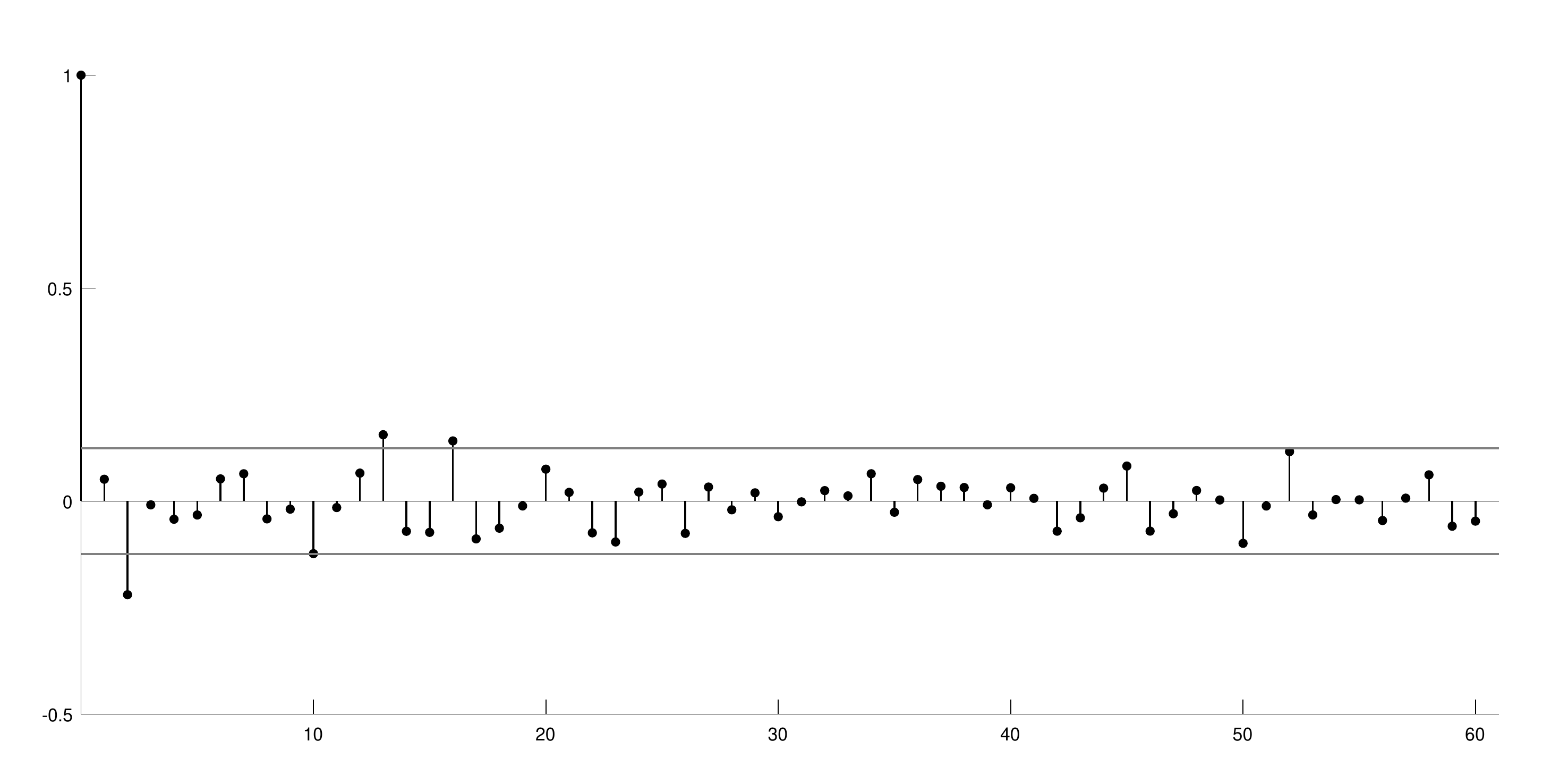}
		\caption{autocorrelation function, weekly, after exposure adjustment}
		\label{fig:nsw_data_w_adj}
	\end{subfigure}	
	
	\begin{subfigure}{0.49\textwidth}
		\centering
		\includegraphics[width=\textwidth]{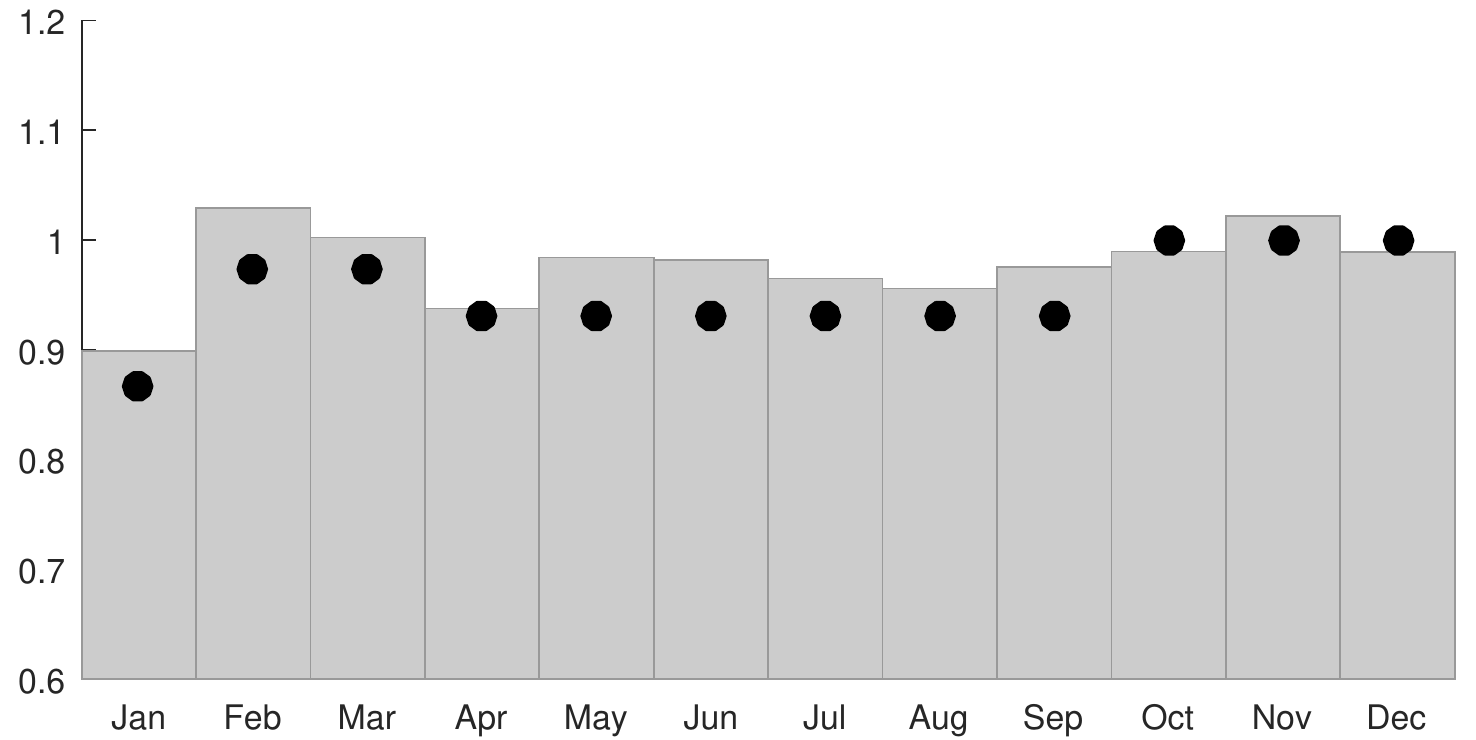}
		\caption{Average claim count per policy - Monthly pattern. \\
			Bar chart - empirical patterns, dark lines - fitted patterns.}
		\label{fig:nsw_moy}
	\end{subfigure} 
	\begin{subfigure}{0.49\textwidth}
		\centering
		\includegraphics[width=\textwidth]{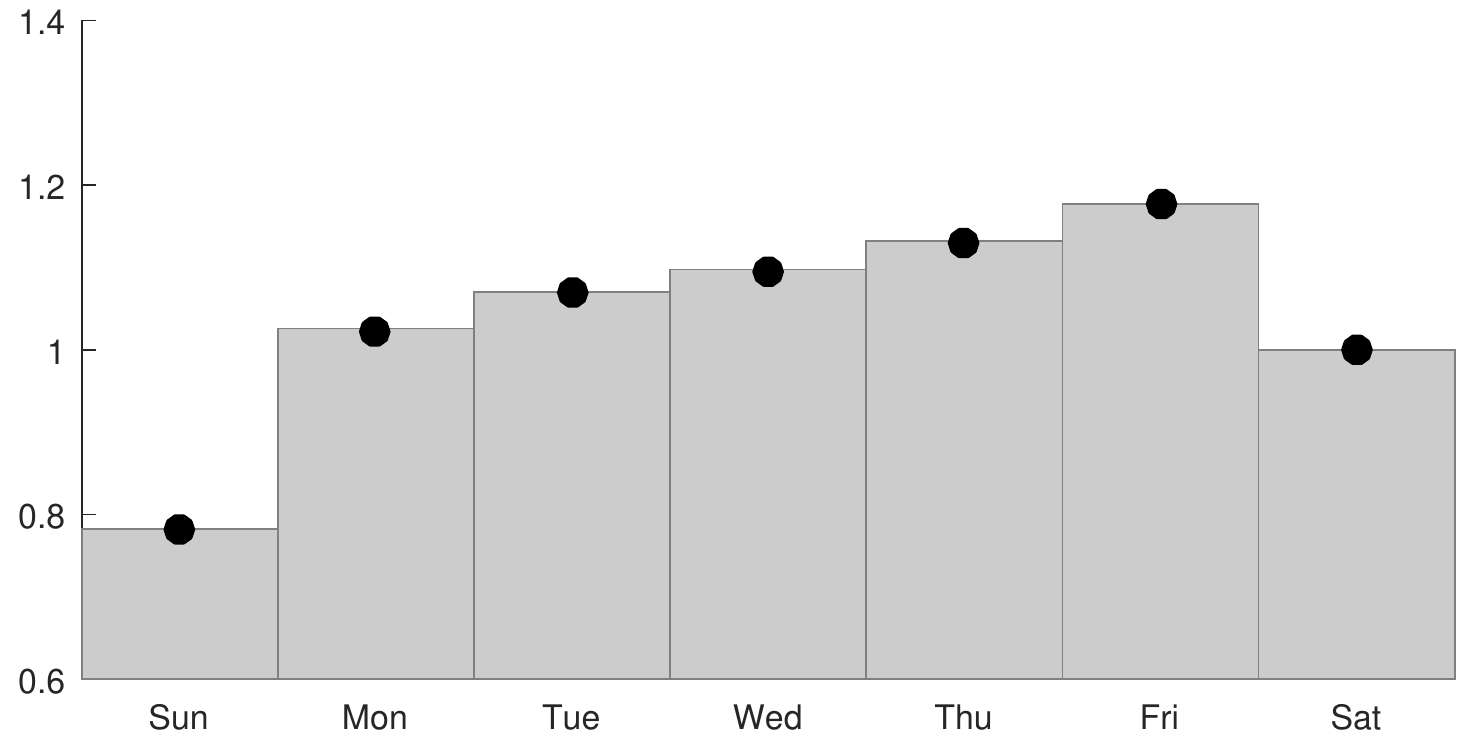}
		\caption{Average claim count per policy - Weekly pattern\\
			Bar chart - empirical patterns, dark dots - fitted patterns.}
		\label{fig:nsw_dow}
	\end{subfigure} 
	
	\caption{Summary of the claim arrival process of the Motor LoB in NSW}
	\label{fig:nsw_count}
\end{figure}

\begin{figure}[!htb]
	\centering	
	
	\begin{subfigure}{\textwidth}
		\centering
		\includegraphics[width=\textwidth]{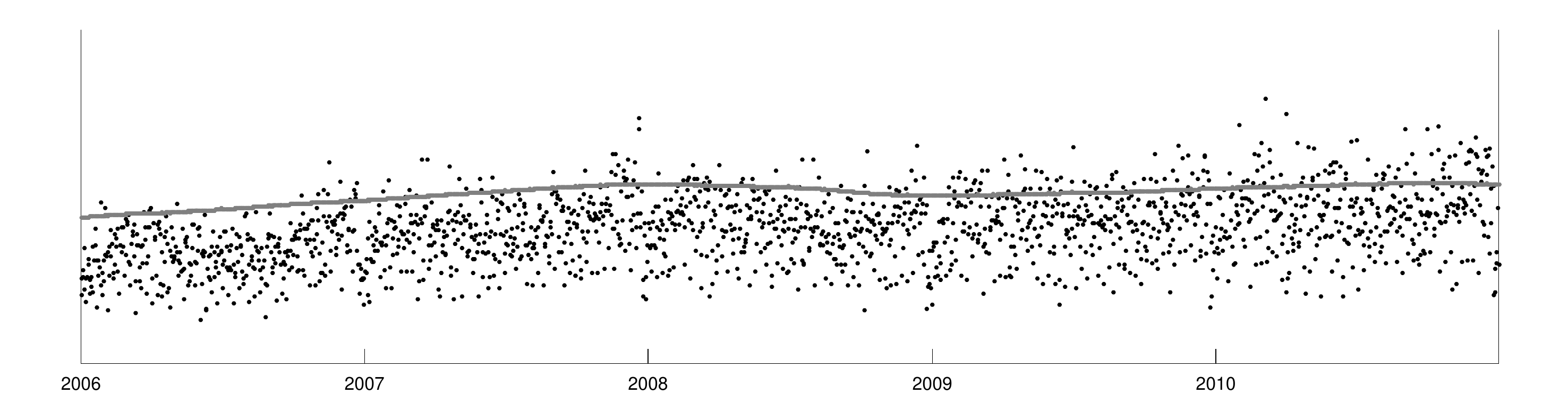}
		\subcaption{Daily claim count (dark) and policy count (grey)}
		\label{fig:vic_data_ts}
	\end{subfigure}
	
	\begin{subfigure}{0.49\textwidth}
		\centering
		\includegraphics[width=\textwidth]{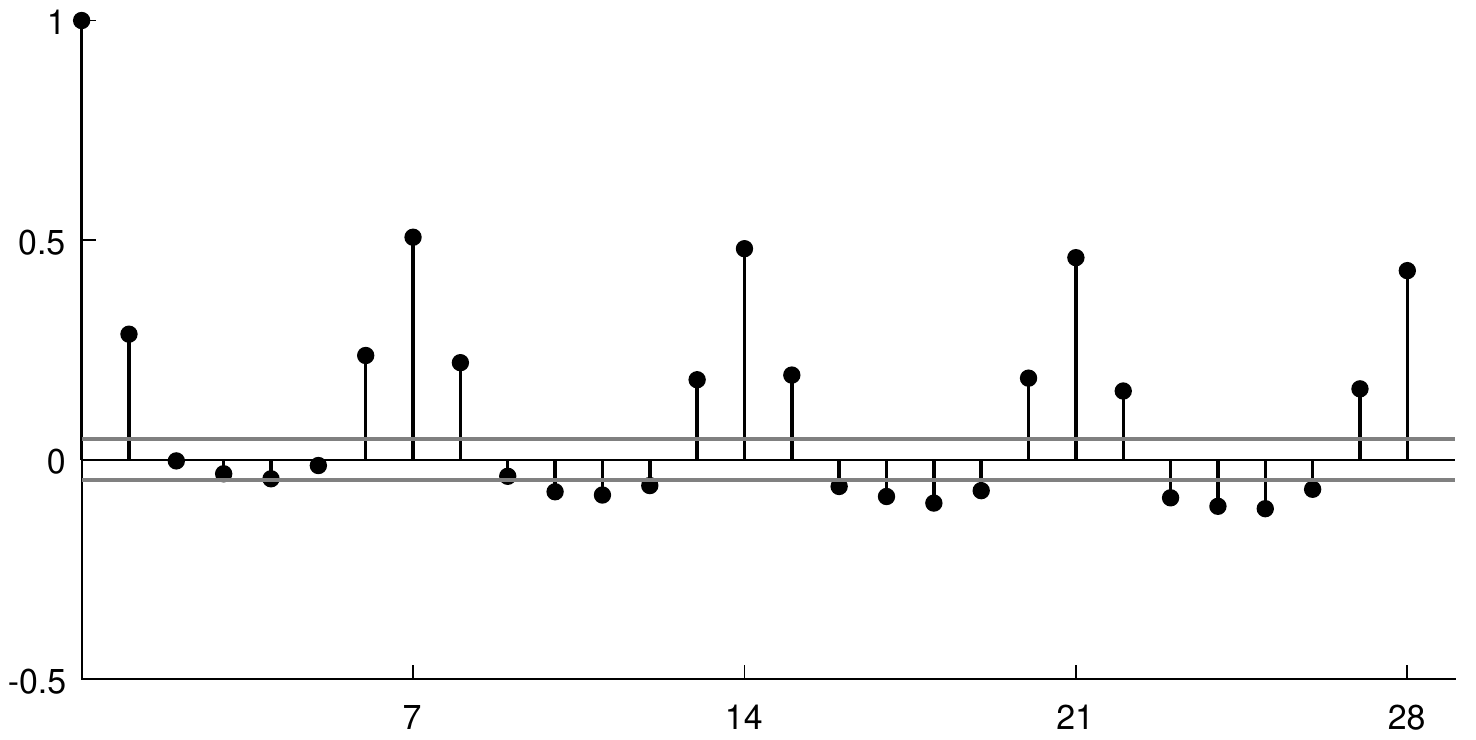}
		\caption{autocorrelation function, daily, raw}
		\label{fig:vic_data_d}
	\end{subfigure}
	\begin{subfigure}{0.49\textwidth}
		\centering
		\includegraphics[width=\textwidth]{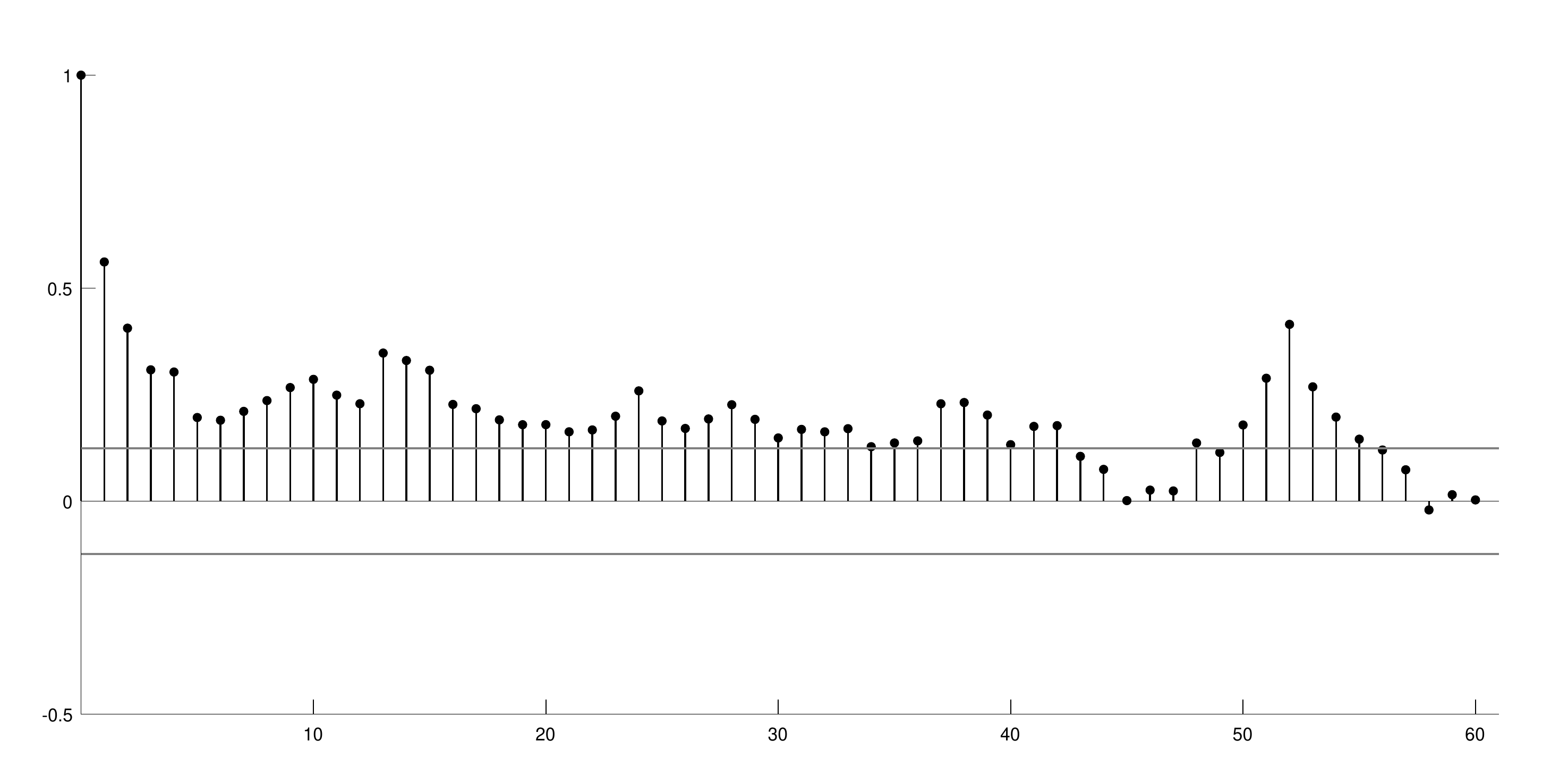}
		\caption{autocorrelation function, weekly, raw}
		\label{fig:vic_data_w}
	\end{subfigure}	

	\begin{subfigure}{0.49\textwidth}
		\centering
		\includegraphics[width=\textwidth]{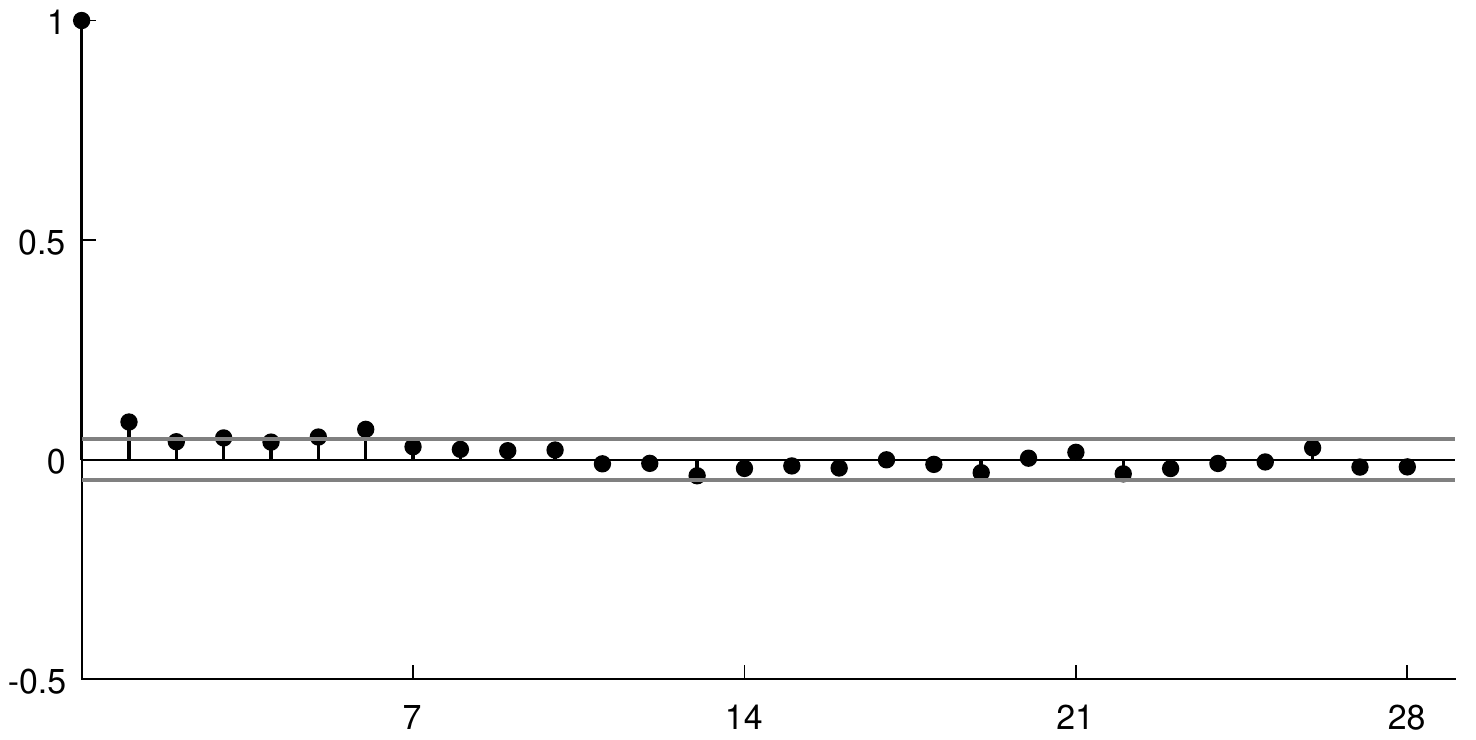}
		\caption{autocorrelation function, daily, after exposure adjustment}
		\label{fig:vic_data_d_adj}
	\end{subfigure}
	\begin{subfigure}{0.49\textwidth}
		\centering
		\includegraphics[width=\textwidth]{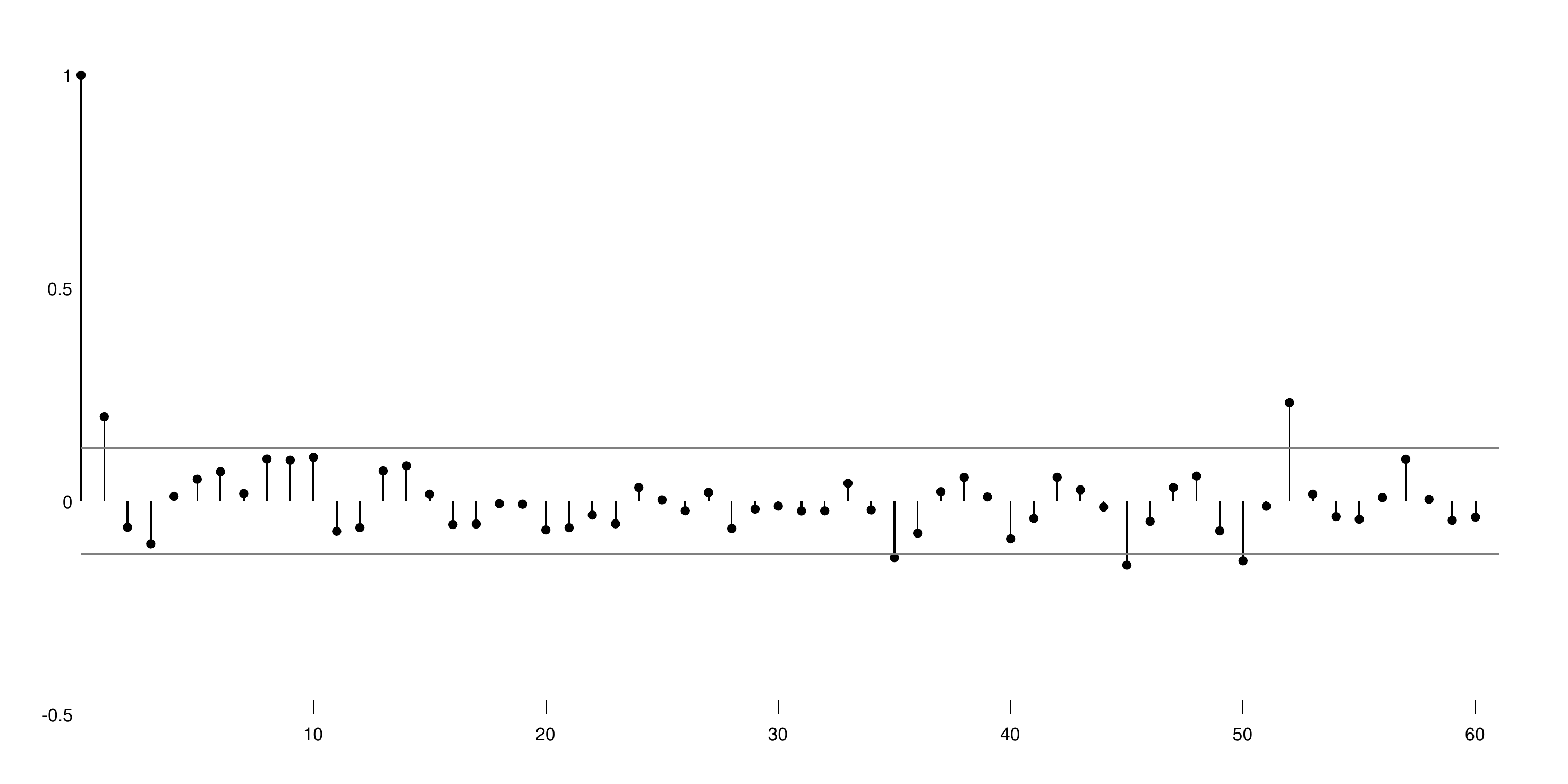}
		\caption{autocorrelation function, weekly, after exposure adjustment}
		\label{fig:vic_data_w_adj}
	\end{subfigure}
	
	\begin{subfigure}{0.49\textwidth}
		\centering
		\includegraphics[width=\textwidth]{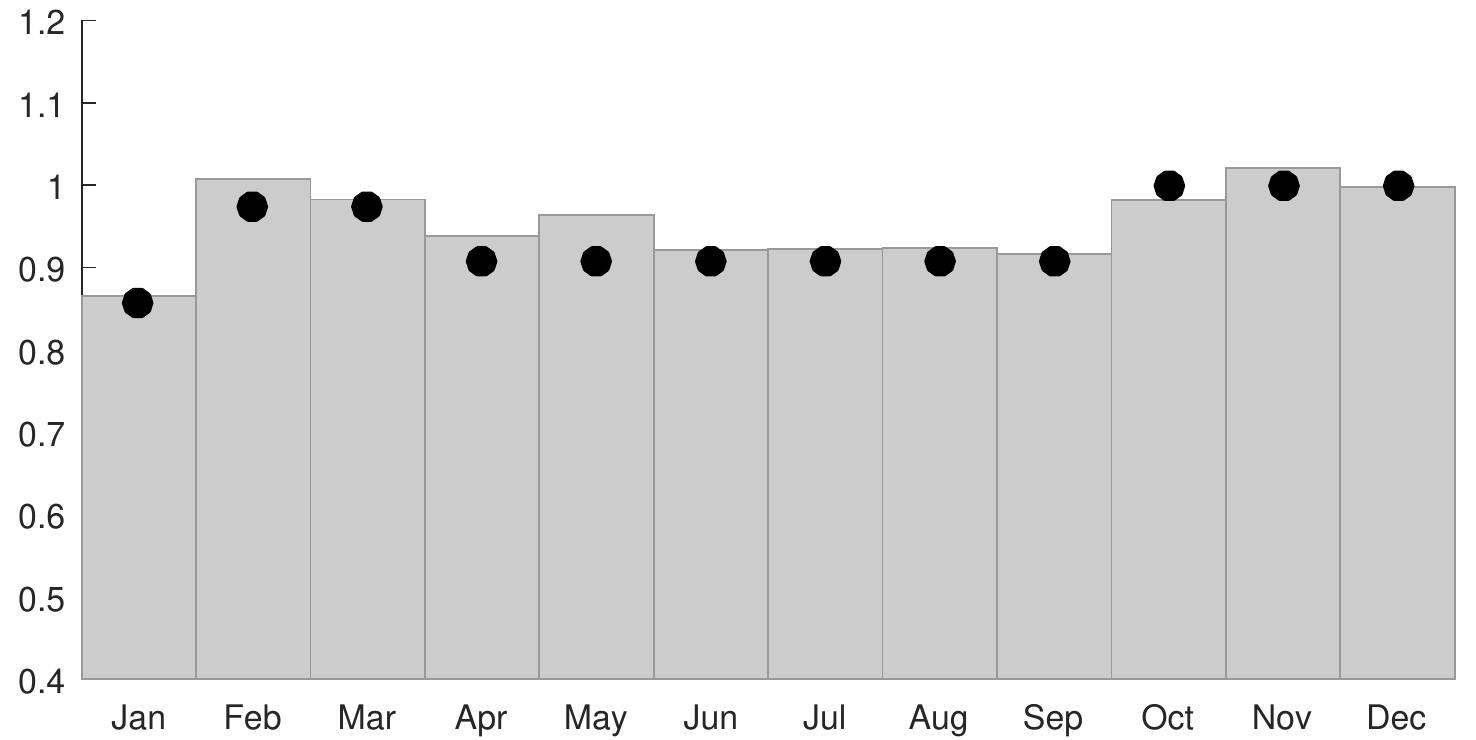}
		\caption{Average claim count per policy - Monthly pattern\\
			Bar chart - empirical patterns, dark lines - fitted patterns.}
		\label{fig:vic_moy}
	\end{subfigure} 
	\begin{subfigure}{0.49\textwidth}
		\centering
		\includegraphics[width=\textwidth]{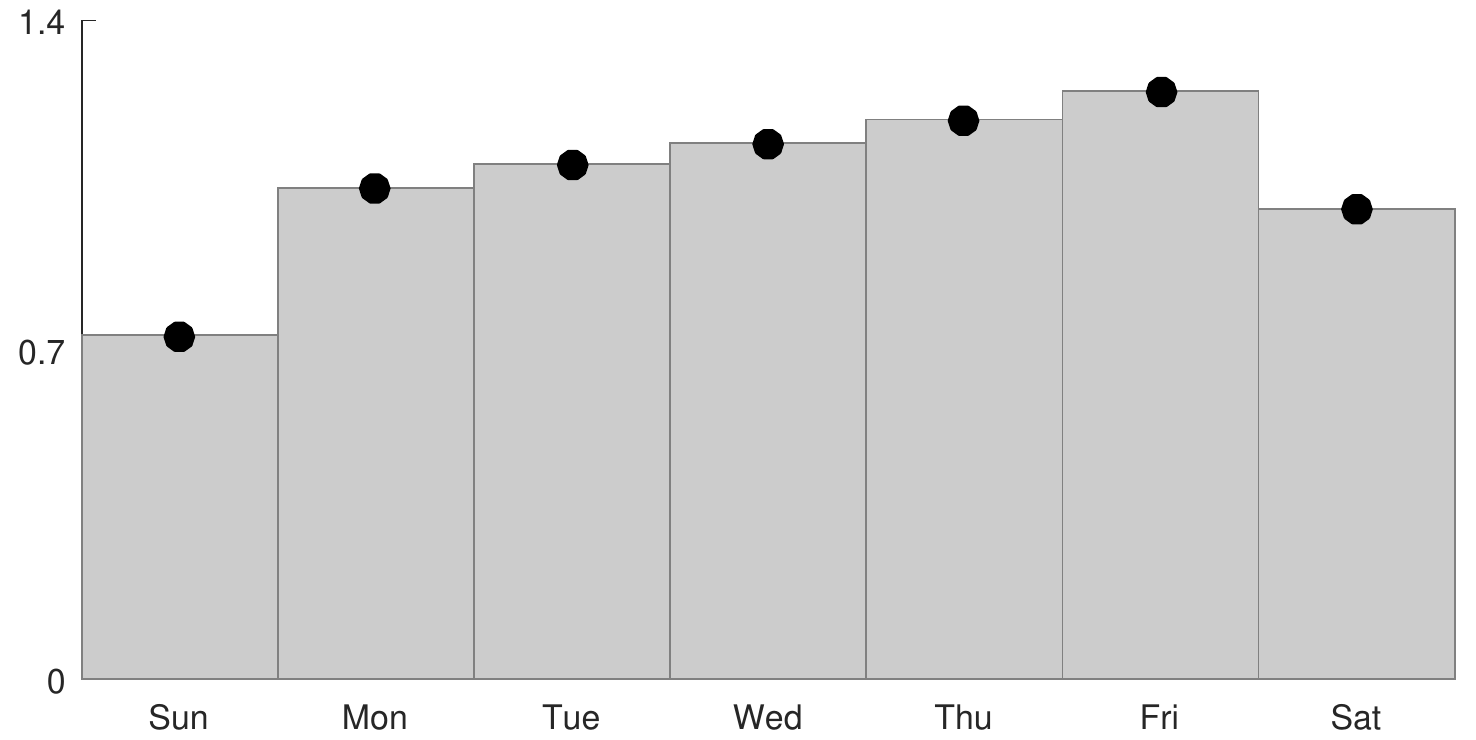}
		\caption{Average claim count per policy - Weekly pattern\\
			Bar chart - empirical patterns, dark dots - fitted patterns.}
		\label{fig:vic_dow}
	\end{subfigure} 
	
	\caption{Summary of the claim arrival process of the Motor LoB in VIC}
	\label{fig:vic_count}
\end{figure}

Figures \ref{fig:nsw_count} and \ref{fig:vic_count} present the time series of daily claim counts, numbers of policy holders, autocorrelation of daily claim counts (standardised by numbers of daily policyholders) as well as autocorrelation of weekly claim counts (standardised by numbers of weekly policyholders) for both states. This is to account for the non-constant risk exposure of total claims counts and to hence reveal a clearer picture of potential trends, weekly and seasonal patterns of claim frequencies. 

There are a few key observations. Firstly, Figures \ref{fig:nsw_data_ts} and \ref{fig:vic_data_ts} present the reported claim count per accident days along with the numbers of policies. The daily autocorrelation plots (see Figures \ref{fig:nsw_data_d} and \ref{fig:vic_data_d}) reveal strong weekly cycles for both states. Furthermore, the weekly autocorrelation plots (see Figures \ref{fig:nsw_data_w} and \ref{fig:vic_data_w}) suggest significant annual cycles for both states.

The existence of strong empirical autocorrelations (see, e.g., Figures \ref{fig:nsw_data_d} and \ref{fig:vic_data_d}) means one should investigate the seasonality patterns from the data. Furthermore, we have also noticed the presence of potential trends in the data. We have carried an empirical investigation where we compare the average claim count per policy across (1) different days of a week, (2) different months and (3) during major Australian holidays. Please see the bar charts in Figures \ref{fig:nsw_moy} and \ref{fig:nsw_dow} the empirical monthly and weekly patterns of (standardised) claim count per policy for NSW, and the bar charts in Figures \ref{fig:vic_moy} and \ref{fig:vic_dow} for VIC.

Let us further specify the risk exposure, $W_g(t)$ (where $g=$ NSW, VIC) as 
\begin{equation}
	\log W_g(t)=\log\left(\text{number of policies in force at time } t\right)+ f_g(t),
\end{equation}
where a deterministic function of time, $f_g(t)$, is adopted to capture time covariates (e.g. seasonality, trends, etc.). The particular form of $f_g(t)$ depends on the features of data. 

In this project, we adopt an additive structure of $f_g(t)$ such that 
\begin{equation}
	f_g(t)=\boldsymbol{x_g}(t)\boldsymbol{a_g}\\
\end{equation}

\noindent where $\boldsymbol{a_g}(t)$ is a vector of covariate coefficients and $\boldsymbol{x_g}(t)$ is a matrix of the covariates (with each row representing each accident day and each column representing each covariate). Note we assume that both risk exposure $W_g(t)$ and covariates $\bo{x_g}(t)$ are piece-wise constant over daily intervals. This is a natural assumption given that daily observations are the most granular level of information one can possibly obtain in practice. 

Joint estimation of the seasonality component of function $f_g$ and the unobservable shot noise component of $\tilde{\lambda}_g(t)$ ($g=$ NSW, VIC) involves a large number of parameters. We propose to estimate the seasonality parameters prior to that of the shot noise parameters. This is achieved by fitting a Poisson GLM model to the daily claim count. This effectively approximates the Cox model by replacing the shot noise component $\tilde{\lambda}_g(t)$ with a constant $c_g$ along with the following discretisation:
\begin{equation}
	\log(\tilde{\lambda}_g(\lfloor t\rfloor+1))=\log(W_g(\lfloor t\rfloor+1))+\boldsymbol{x_g(\lfloor t\rfloor+1)}\boldsymbol{a_g}+c_g.
\end{equation}
where $\lfloor t\rfloor$ is the integer part of $t$ (with unit of day). Such a method effectively fits a Poisson GLM model with a log-link function to the data of daily claim counts. 

There can be potentially a large number of parameters in the specification of the seasonality components --- in particular, for the monthly patterns. We have attempted to reduce the number of monthly covariates in two different ways. The first method applies a linear spline model which is aimed at capturing the change of monthly behaviour of claim frequencies with a simple and continuous parametric form. The second method involves grouping months (hence requires a benchmark month). Both methods were used and we selected the one that performed better based on the AIC criterion. It turns out that the method of grouping works better for both the claim processes of NSW and VIC. We have also attempted to reduce the number of weekly covariates by grouping days in a week, however, it turns out that having 6 parameters for the weekly pattern outperforms grouping (in terms of the AIC criteria) for both states. The results of fitting the monthly and weekly patterns are presented in the dark dots in Figures \ref{fig:nsw_moy} and \ref{fig:vic_moy}, as well as in dark dots in Figures \ref{fig:nsw_dow} and \ref{fig:vic_dow}. 

We adjust the empirical autocorrelation plots by examining daily claim count per exposure. The empirical autocorrelations after adjustment are plotted in Figures \ref{fig:nsw_data_d_adj}, \ref{fig:nsw_data_w_adj}, \ref{fig:vic_data_d_adj} and \ref{fig:vic_data_w_adj}. Compared to those before adjustments, both states display significantly reduced levels of autocorrelation. \rev{This suggests that we have modelled away most of what could be explained with our covariates, as would be expected for any appropriate model, and we are now ready to move to the next stage and fit a dependence structure \citep[see][for a detailed discussions of the benefits and requirements of modelling trends before applying a dependence structure]{AvTaWo16}.}

\subsection{Univariate claim arrival analysis}\label{sec:fitting_univariate}

\rev{The fitting of L\'evy copulas to `assemble' processes separate from the marginal processes. Hence, we start by modelling our margins.} The calibration of each univariate Cox process follows a similar procedure as \citet*{AvWoYa16}, which includes the following steps. 

The initial estimates for the shot noise parameters are obtained by matching of moments, followed by jointly updating both the reporting delay and shot noise parameters through MCEM algorithms. This involves 150 MCEM iterations with 20,000 RJMCMC simulations; and we select 100 simulations from the second half of each iteration in evaluating the M-step of the MCEM algorithm.

\begin{figure}[htb]
	\centering
	\begin{subfigure}{0.3\textwidth}
		\centering
		\includegraphics[width=\textwidth]{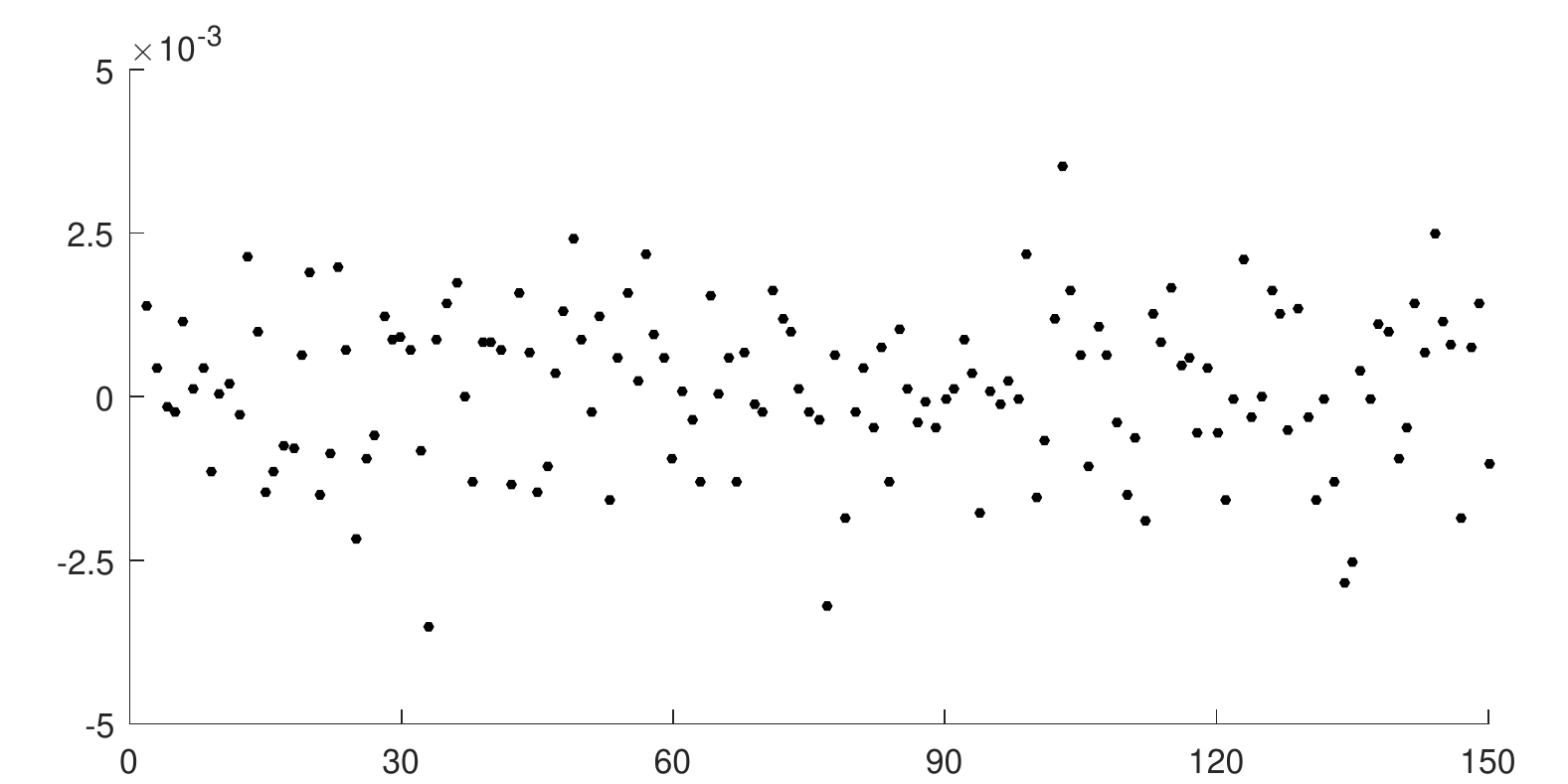}
		\caption{$\rho_{\text{nsw}}$}
	\end{subfigure} 
	\begin{subfigure}{0.3\textwidth}
		\centering
		\includegraphics[width=\textwidth]{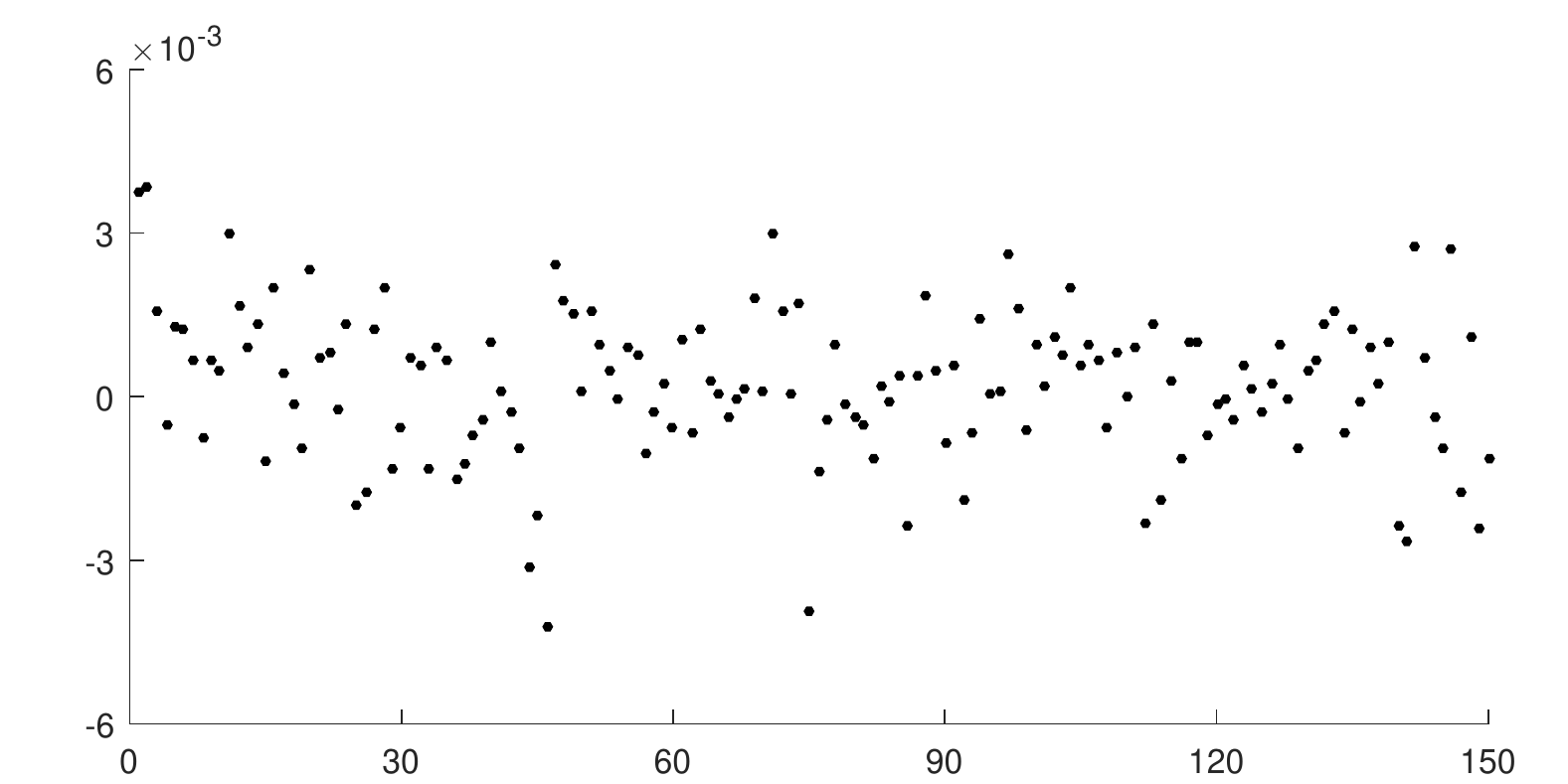}
		\caption{$\eta_{\text{nsw}}$}
	\end{subfigure} 
	\begin{subfigure}{0.3\textwidth}
		\centering
		\includegraphics[width=\textwidth]{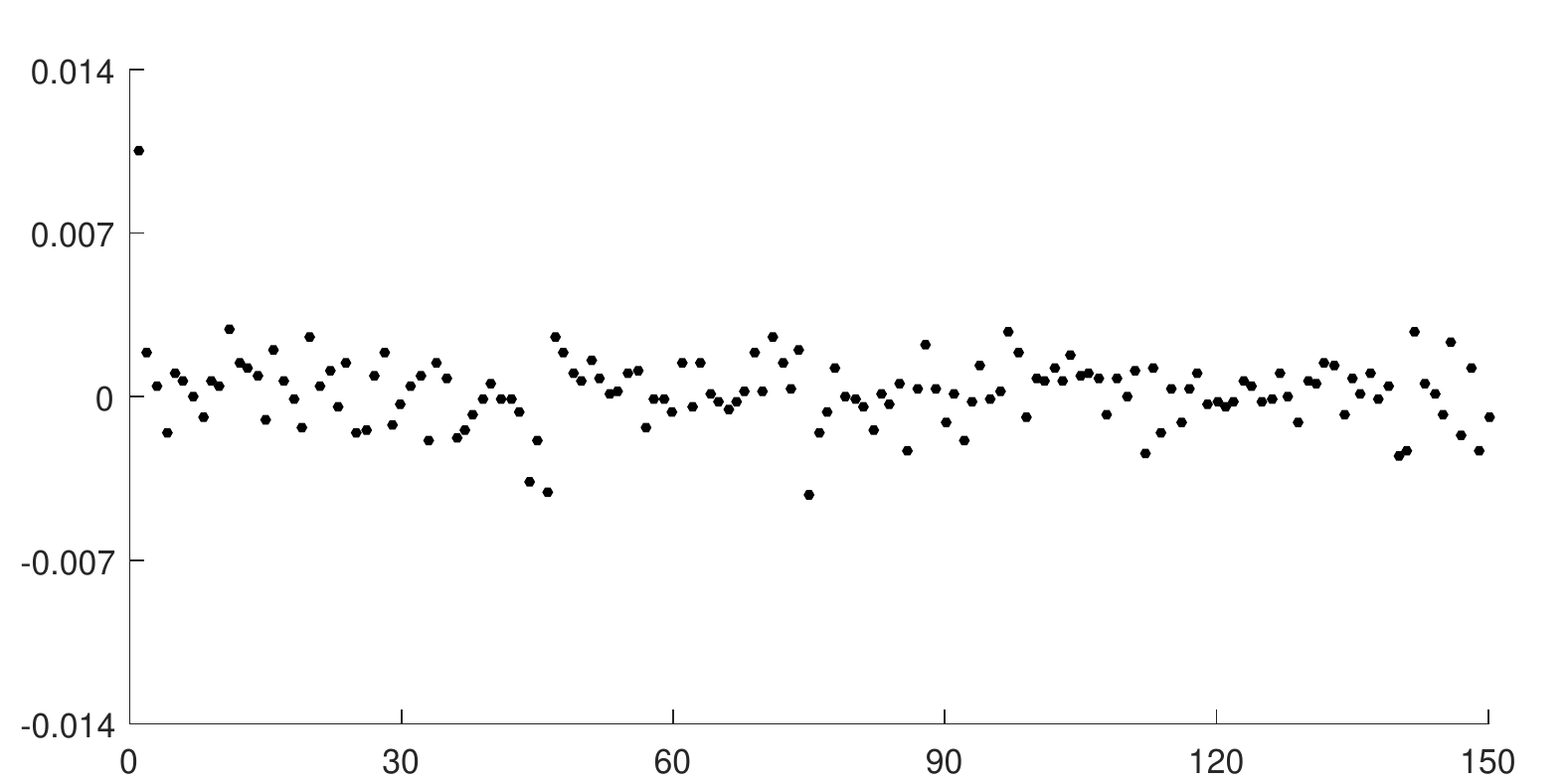}
		\caption{$\rho_{\text{nsw}}/k_{\text{nsw}}$}
	\end{subfigure}

	\begin{subfigure}{0.3\textwidth}
		\centering
		\includegraphics[width=\textwidth]{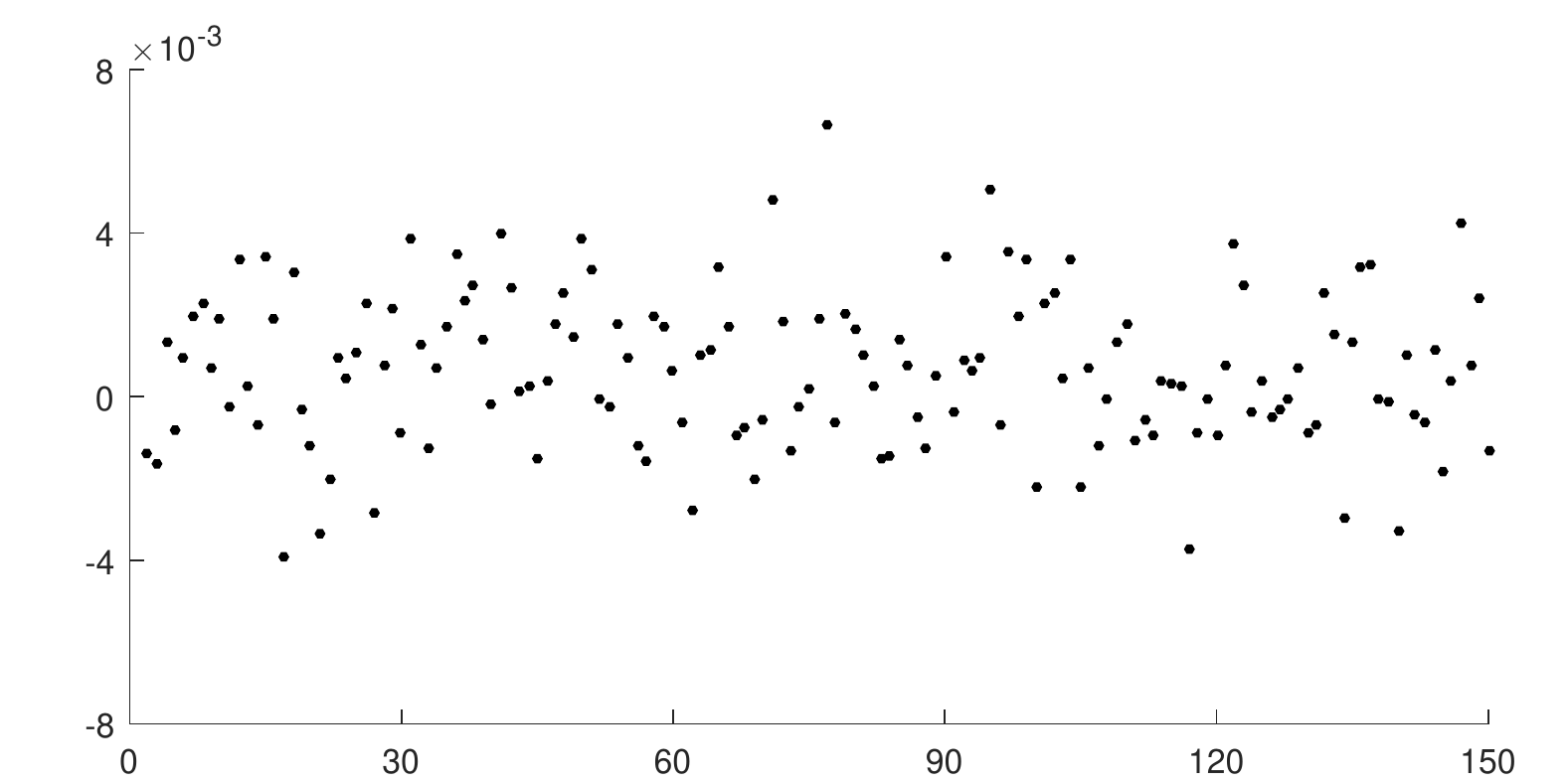}
		\caption{$\rho_{\text{vic}}$}
	\end{subfigure} 
	\begin{subfigure}{0.3\textwidth}
		\centering
		\includegraphics[width=\textwidth]{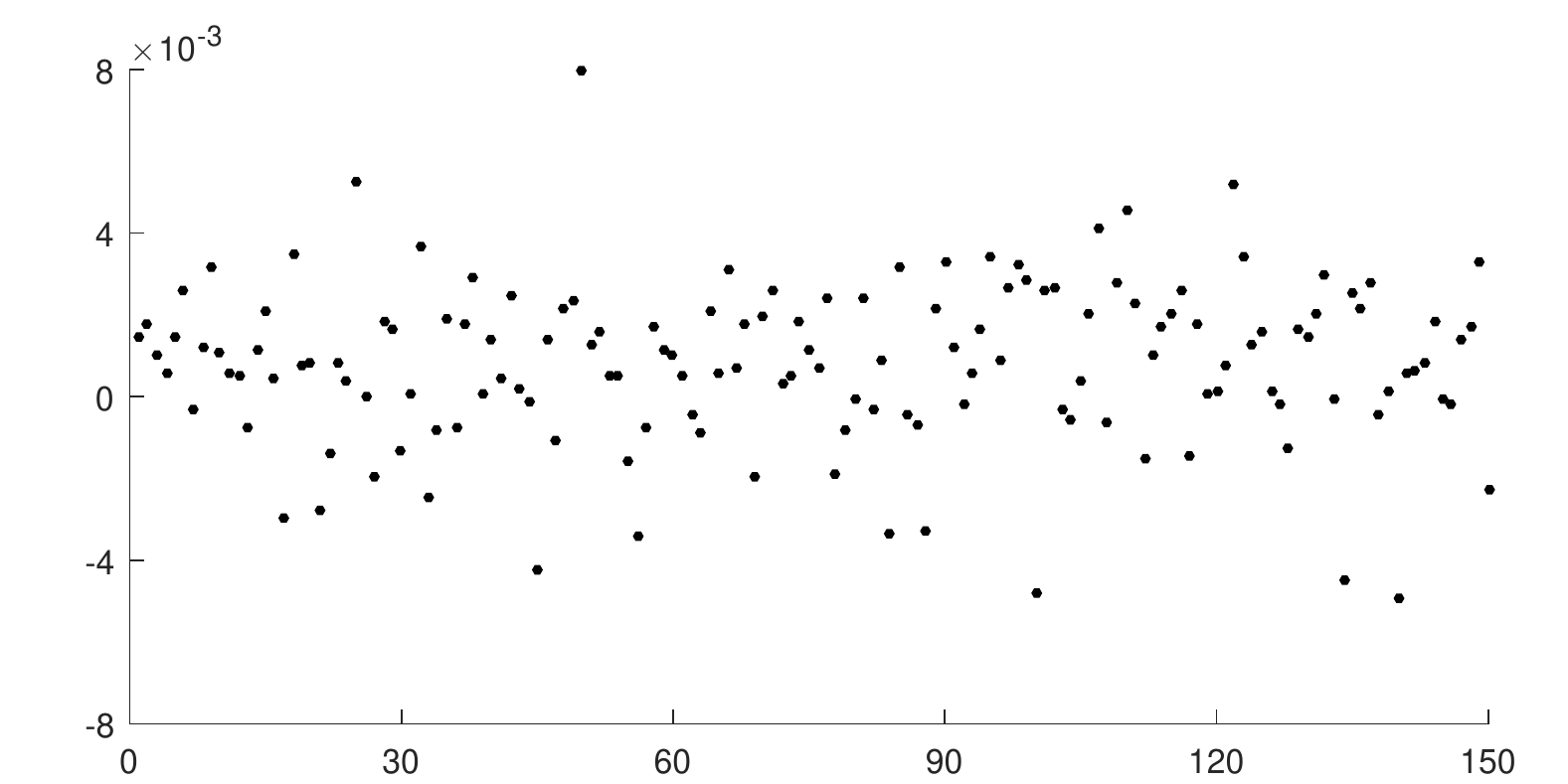}
		\caption{$\eta_{\text{vic}}$}
	\end{subfigure} 
	\begin{subfigure}{0.3\textwidth}
		\centering
		\includegraphics[width=\textwidth]{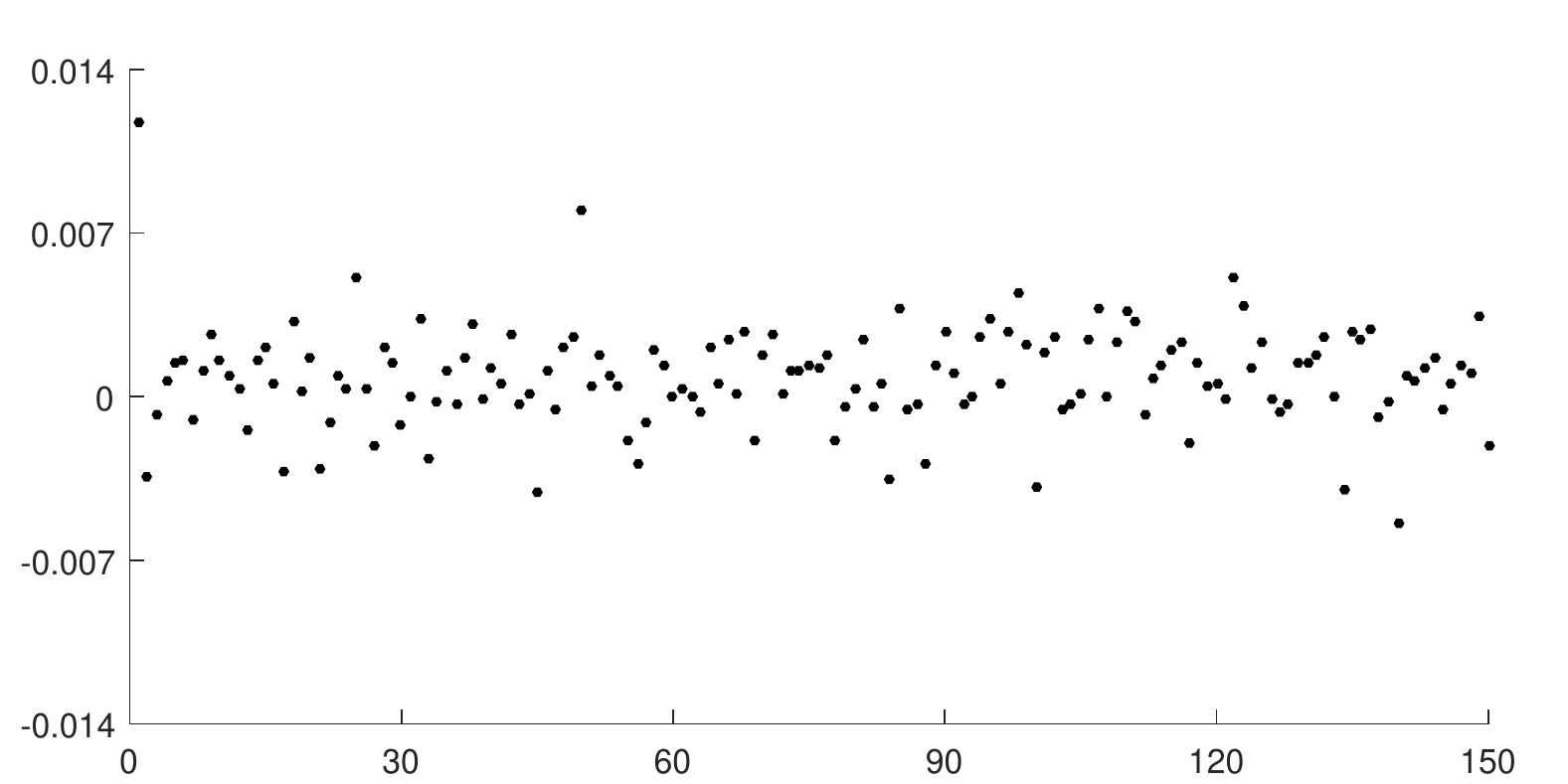}
		\caption{$\rho_{\text{vic}}/k_{\text{vic}}$}
	\end{subfigure}
	\caption{Relative changes of parameter estimates through EM iterations. \rev{Top} - NSW, \rev{Botton} - VIC}
	\label{fig:sn_em}
\end{figure}

The final parameter estimates are summarised in Table \ref{tab:count_param}. Here parameters $\rho$, $\eta$ and $k$ fully specify the marginal shot noise processes (see Theorem \ref{definition:commonshock}), and the implied moments are presented in Table \ref{tab:count_moments}. Figure \ref{fig:sn_em} presents the relative change of parameter estimates (via EM iterations), which shows that the EM estimates are quite stable for all the parameters.  

\begin{table}[htb]
	\centering
	\begin{tabular}{l|lll}
		\hline
		States & $\rho$  & $\eta$ & $\kappa$    \\ \hline
		NSW    & 33.77 & 0.17 & 2.37 \\
		VIC    & 18.74 & 0.18 & 1.28 \\ \hline
	\end{tabular}
	\caption{Parameter estimates of the univariate shot noise Cox processes}
	\label{tab:count_param}
\end{table}

\begin{table}[htb]
	\centering
	\begin{tabular}{llll}
		\hline
		States & mean    & variance & autocorrelation at 1 day lag \\ \hline
		NSW    & 85.22 & 350.78 & 0.21                        \\
		VIC    & 80.74 & 384.33 & 0.37                        \\ \hline
	\end{tabular}
	\caption{Implied moments of claim count per accident day per person based on estimated parameters}
	\label{tab:count_moments}
\end{table}

\begin{figure}[htb]
	\centering	
	\begin{subfigure}{0.45\textwidth}
		\includegraphics[width=\textwidth]{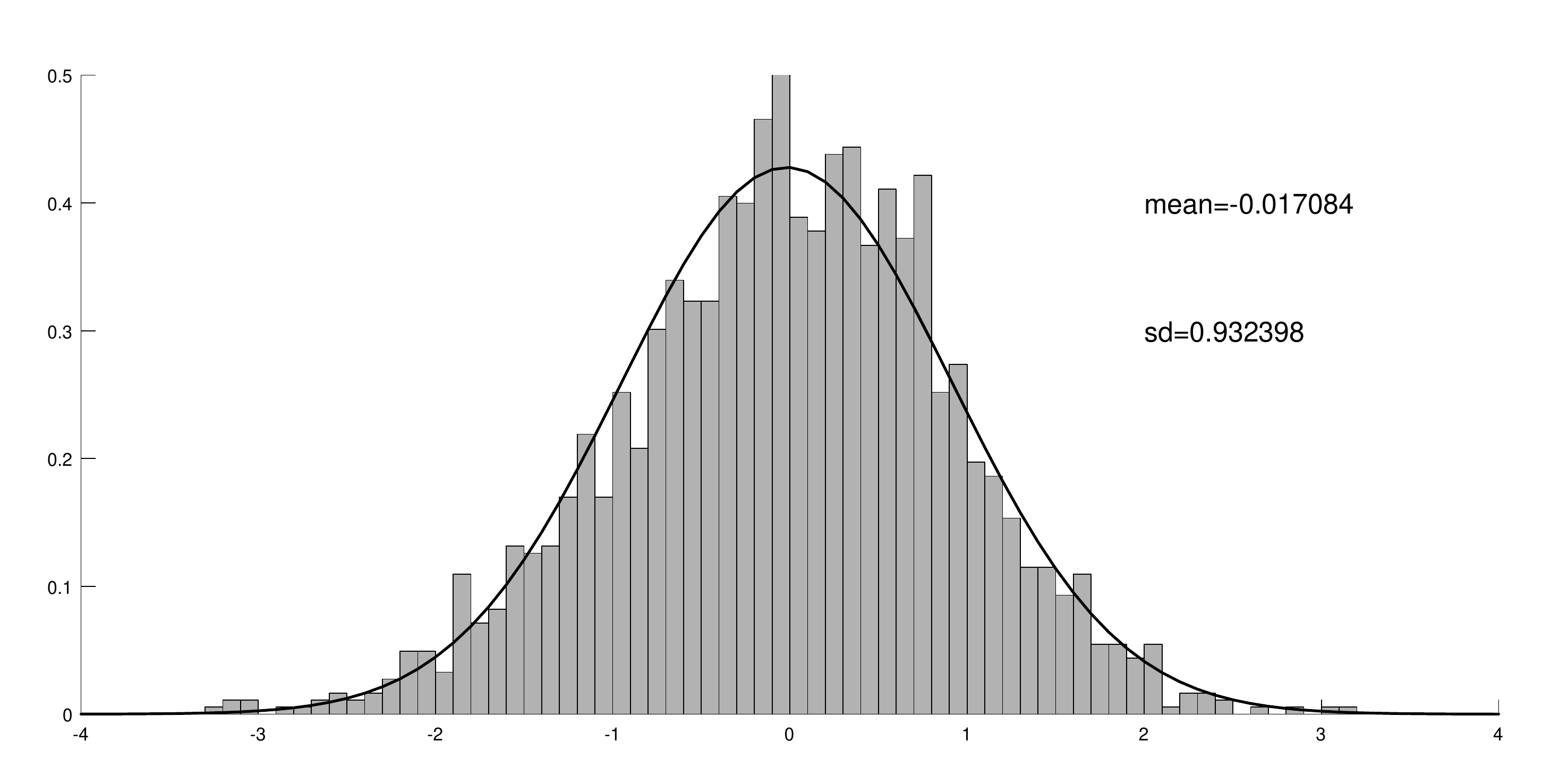}
		\caption{Histograms of residuals and the fitted density with a normal distribution - NSW}
		\label{fig:nsw_residuals_hist}
	\end{subfigure}
	\begin{subfigure}{0.45\textwidth}
		\includegraphics[width=\textwidth]{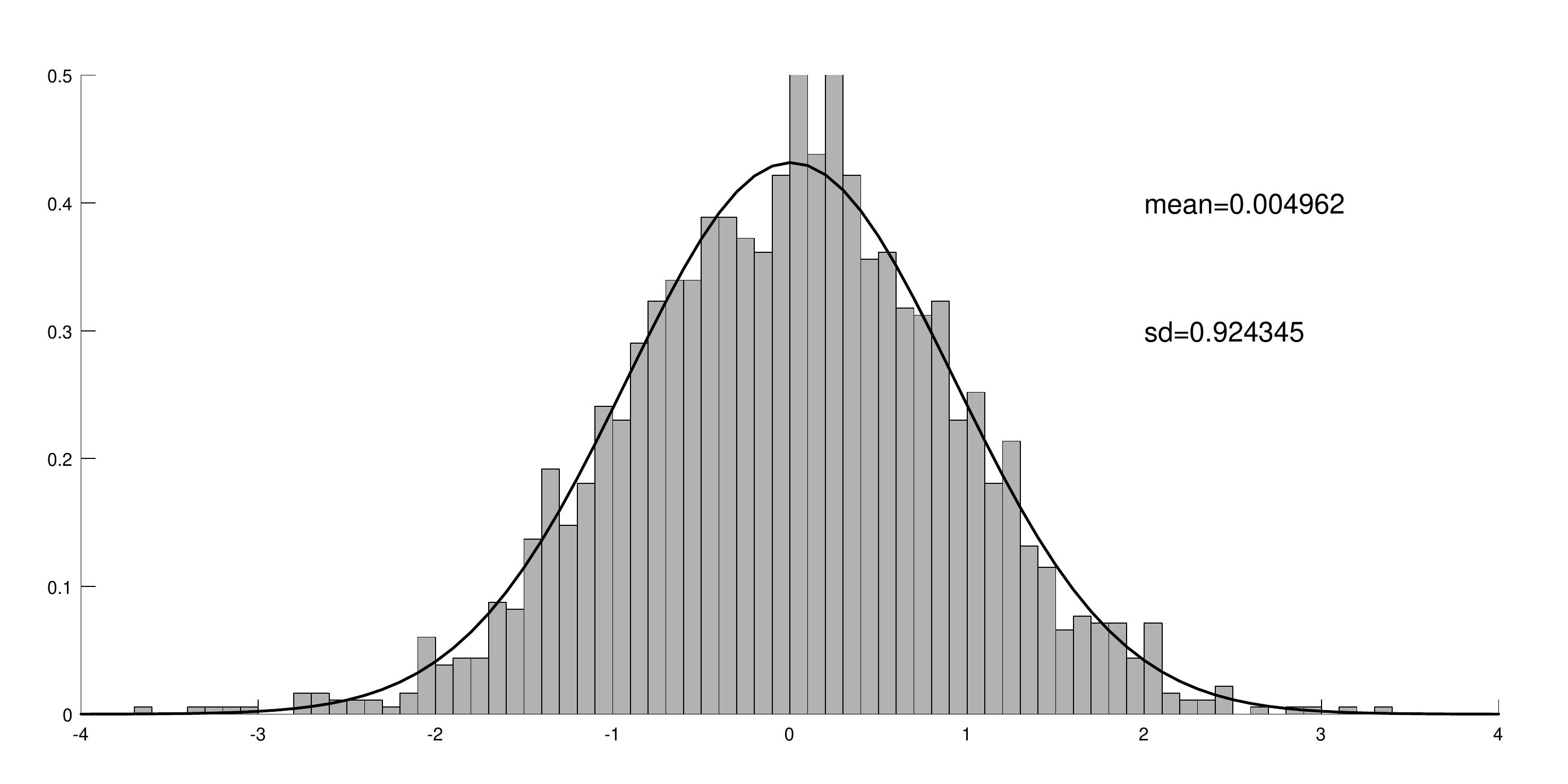}
		\caption{Histograms of residuals and the fitted density with a normal distribution - VIC}
		\label{fig:vic_residuals_hist}
	\end{subfigure}	
	
	\begin{subfigure}{0.45\textwidth}
		\includegraphics[width=\textwidth]{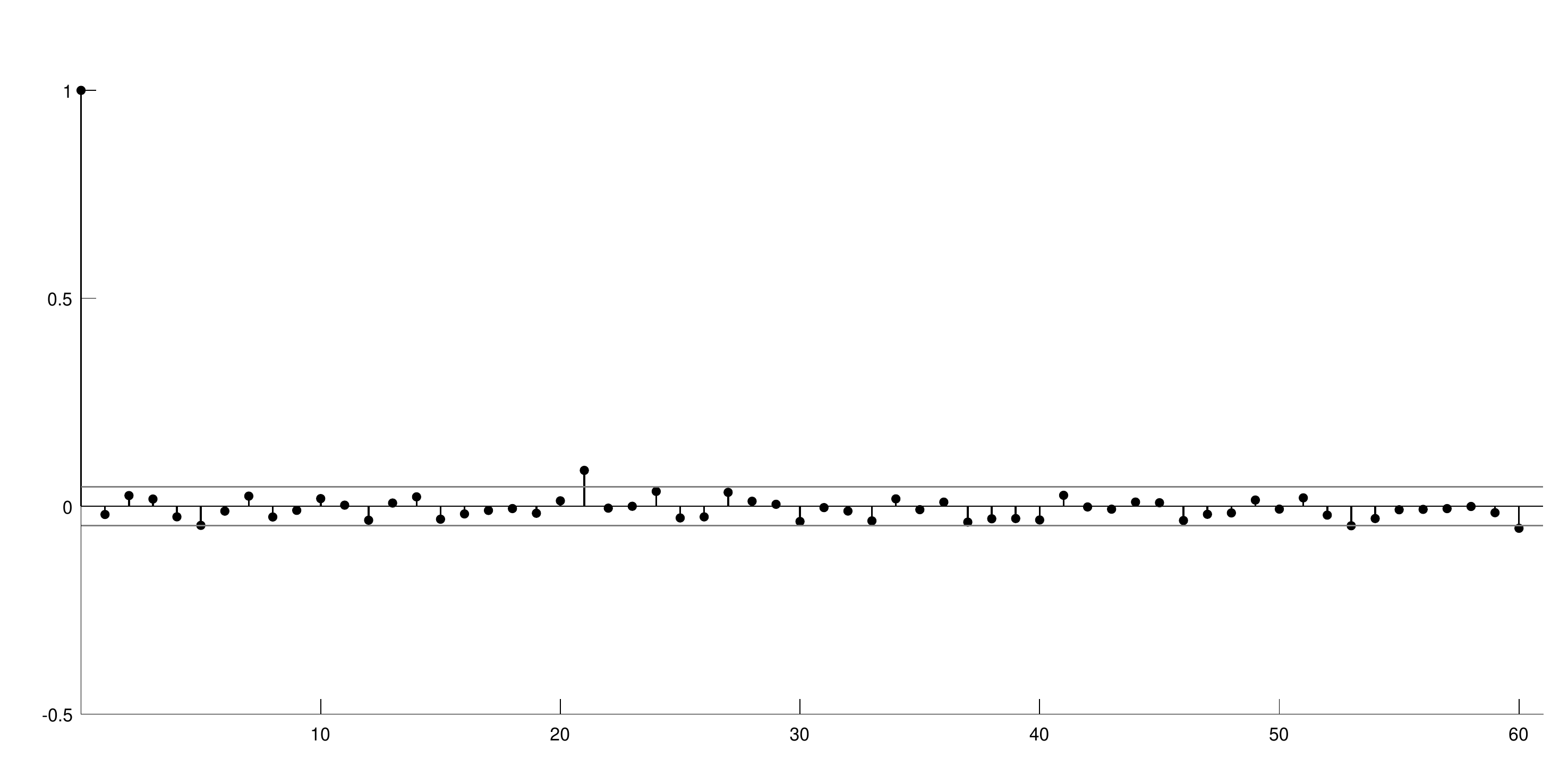}
		\caption{Autocorrelation of residuals - NSW}
		\label{fig:nsw_r_auto_d}
	\end{subfigure}
	\begin{subfigure}{0.45\textwidth}
		\includegraphics[width=\textwidth]{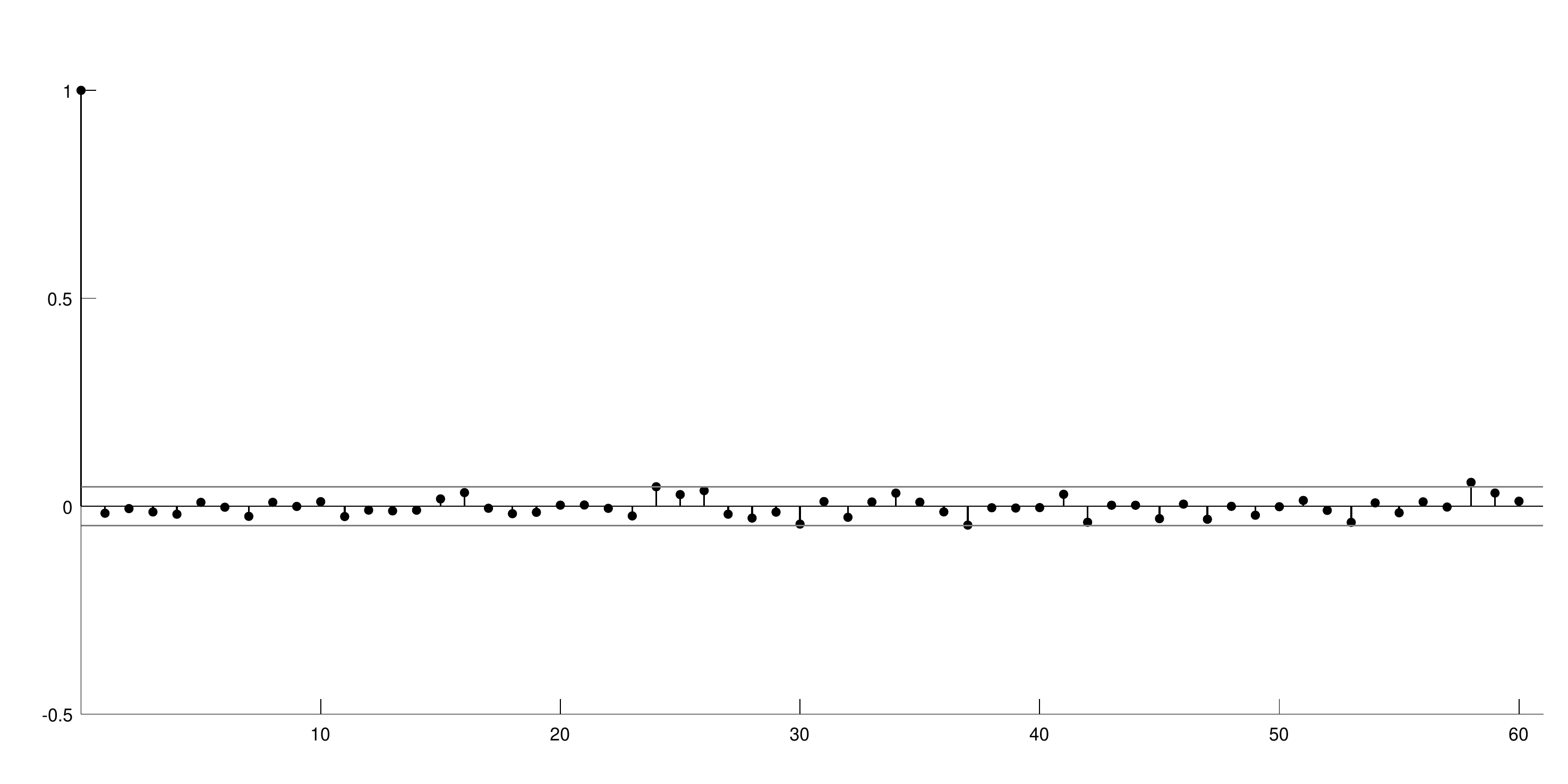}
		\caption{Autocorrelation of residuals - VIC}
		\label{fig:vic_r_auto_d}
	\end{subfigure}	
	\caption{Residual analysis of the independent shot noise Cox process fitting}
	\label{fig:gof_residuals}
\end{figure}

Figures \ref{fig:nsw_residuals_hist} and \ref{fig:vic_residuals_hist} present the assessment of goodness-of-fit via analysing the standardised residuals of estimating daily claim counts (assuming that a claim count is Poisson distributed given the filtered intensity). For both states, the standard deviations of residuals are close to 1 and the means are close to 0, which indicates a satisfactory level of goodness-of-fit. Furthermore, the autocorrelations of the residuals are also close to 0 for both states, which shows that the shot noise Cox model is able to capture the serial dependency of claims counts. 

\subsection{Multivariate claim arrival analysis}\label{sec:fitting_bivariate}

\begin{figure}[!ht]
	\centering	
	\includegraphics[width=0.4\textwidth]{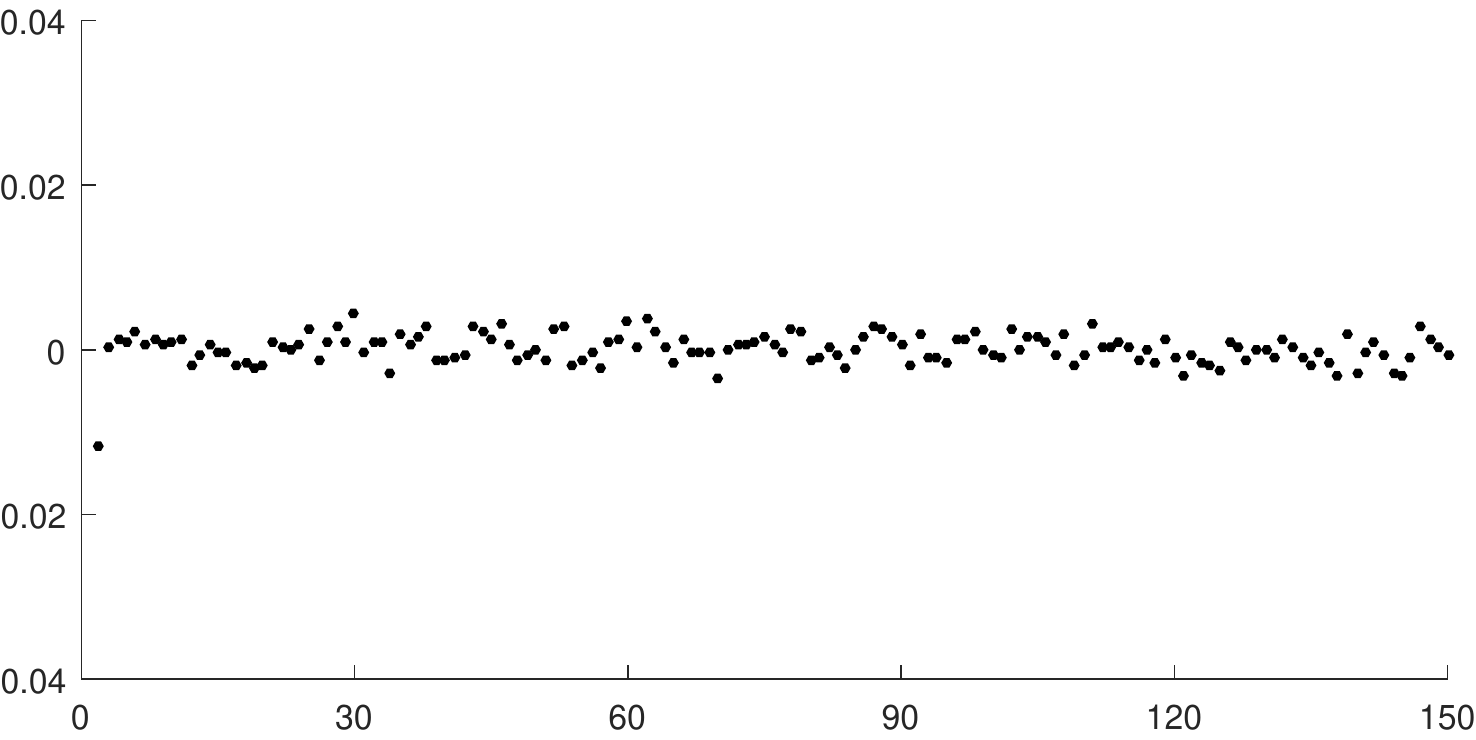}
	\caption{Relative changes of the estimate of the L\'evy copula parameter}
	\label{fig:bi_lc}
\end{figure}

\rev{The final step is to fit a dependence structure between the marginal processes using a L\'evy copula.} In this illustration, we use a Clayton L\'evy copula. The L\'evy copula parameter is updated via 150 MCEM iterations while the marginal parameters of both NSW and VIC are fixed. In each iteration, there are 20,000 RJMCMC simulations and we select 100 simulations from the second half of each iteration in evaluating the M-step of the MCEM algorithm. The final estimate is 0.4214 and the relative change of estimate in each EM iteration is shown in Figure \ref{fig:bi_lc}. 

\begin{figure}[!ht]
	\centering		
	\begin{subfigure}{0.45\textwidth}
		\centering
		\includegraphics[width=\textwidth]{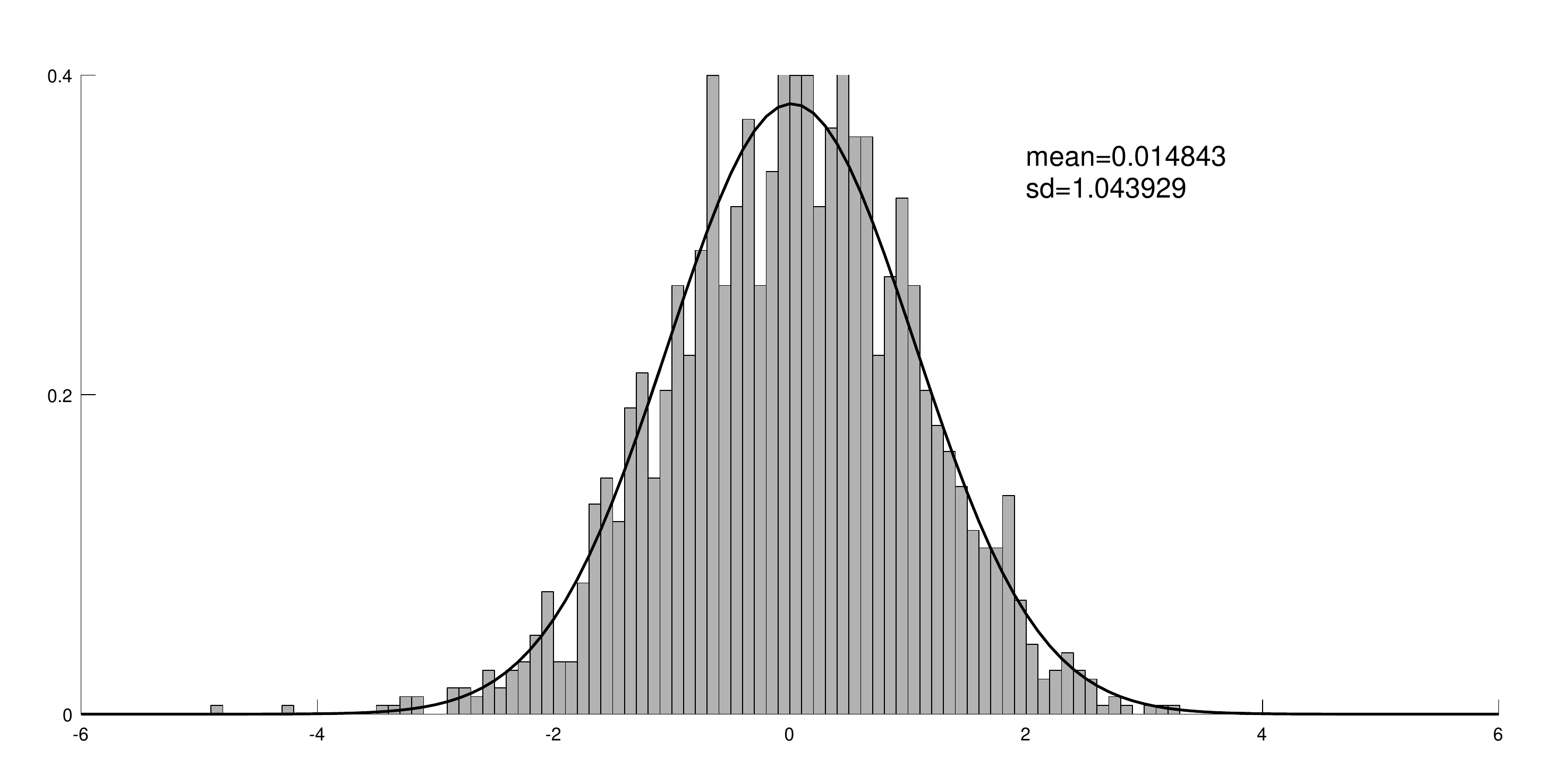}
		\subcaption{Histograms of residuals and the fitted density with a normal distribution - NSW}
		\label{fig:nsw_residuals_bi_hist}
	\end{subfigure}	
	\begin{subfigure}{0.45\textwidth}
		\centering
		\includegraphics[width=\textwidth]{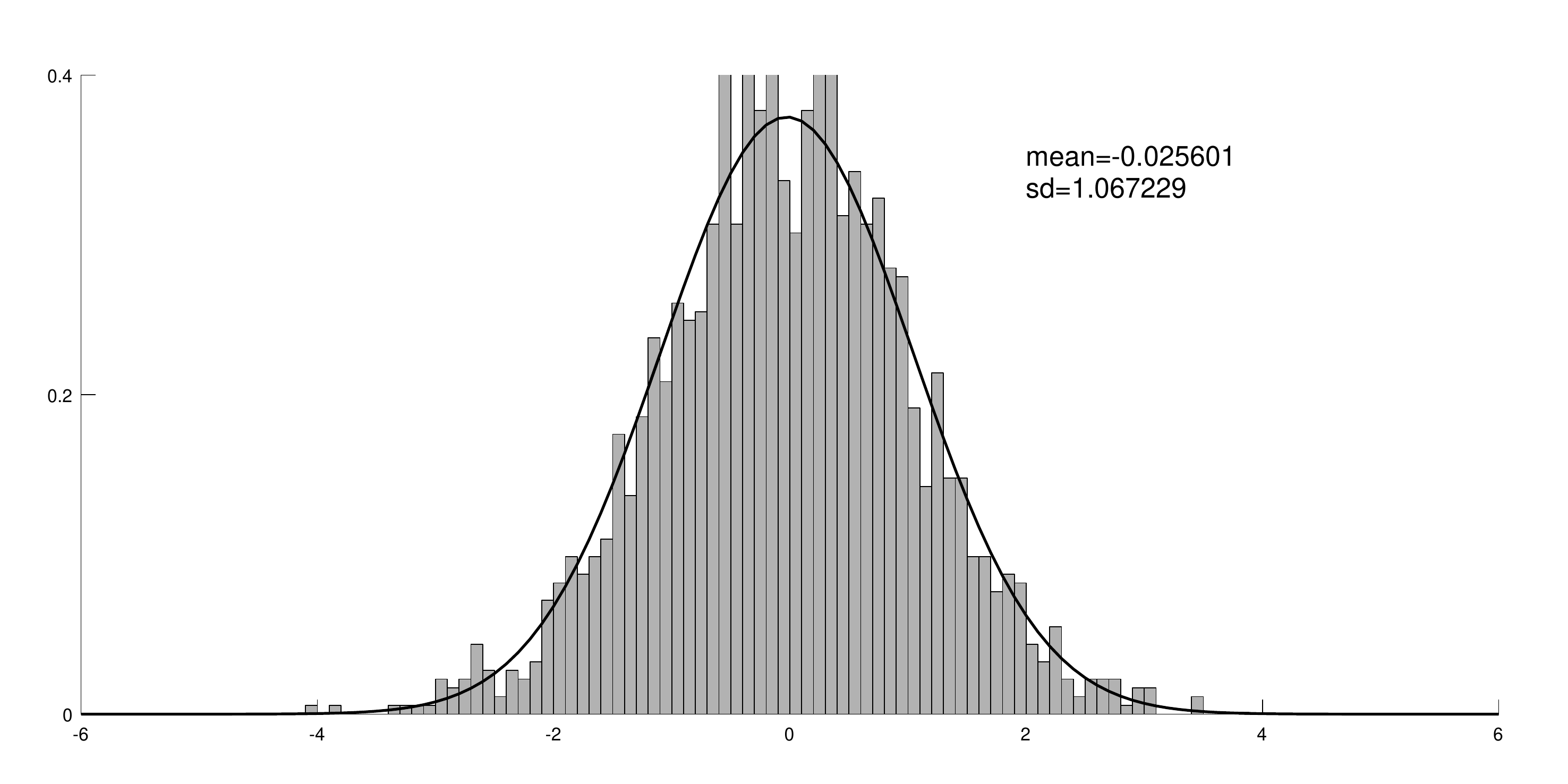}
		\subcaption{Histograms of residuals and the fitted density with a normal distribution - VIC}
		\label{fig:vic_residuals_bi_hist}
	\end{subfigure}

	\begin{subfigure}{0.45\textwidth}
		\centering
		\includegraphics[width=\textwidth]{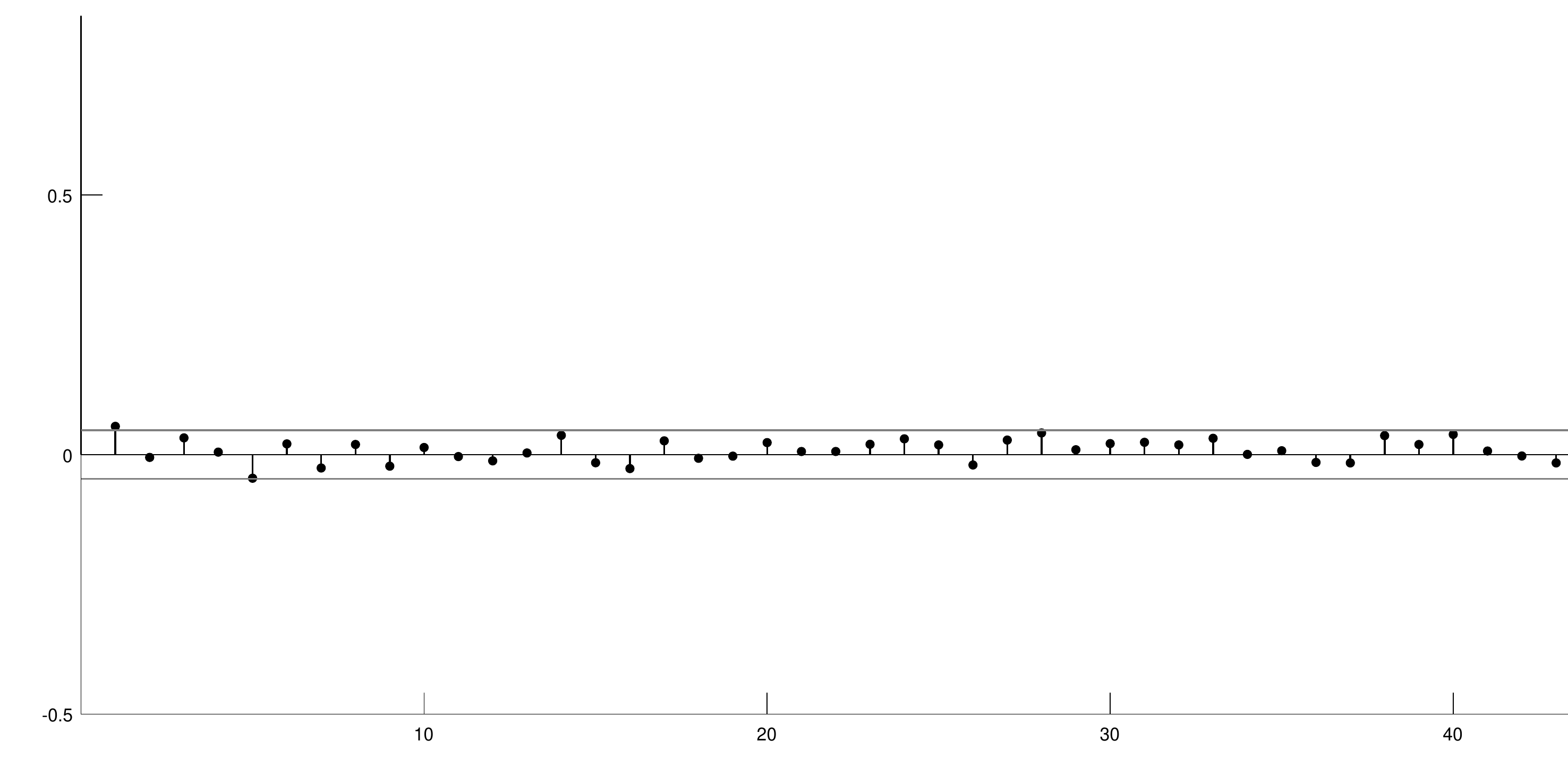}
		\subcaption{Autocorrelation of residuals - NSW}
		\label{fig:nsw_r_bi_auto_d}
	\end{subfigure}	
	\begin{subfigure}{0.45\textwidth}
		\centering
		\includegraphics[width=\textwidth]{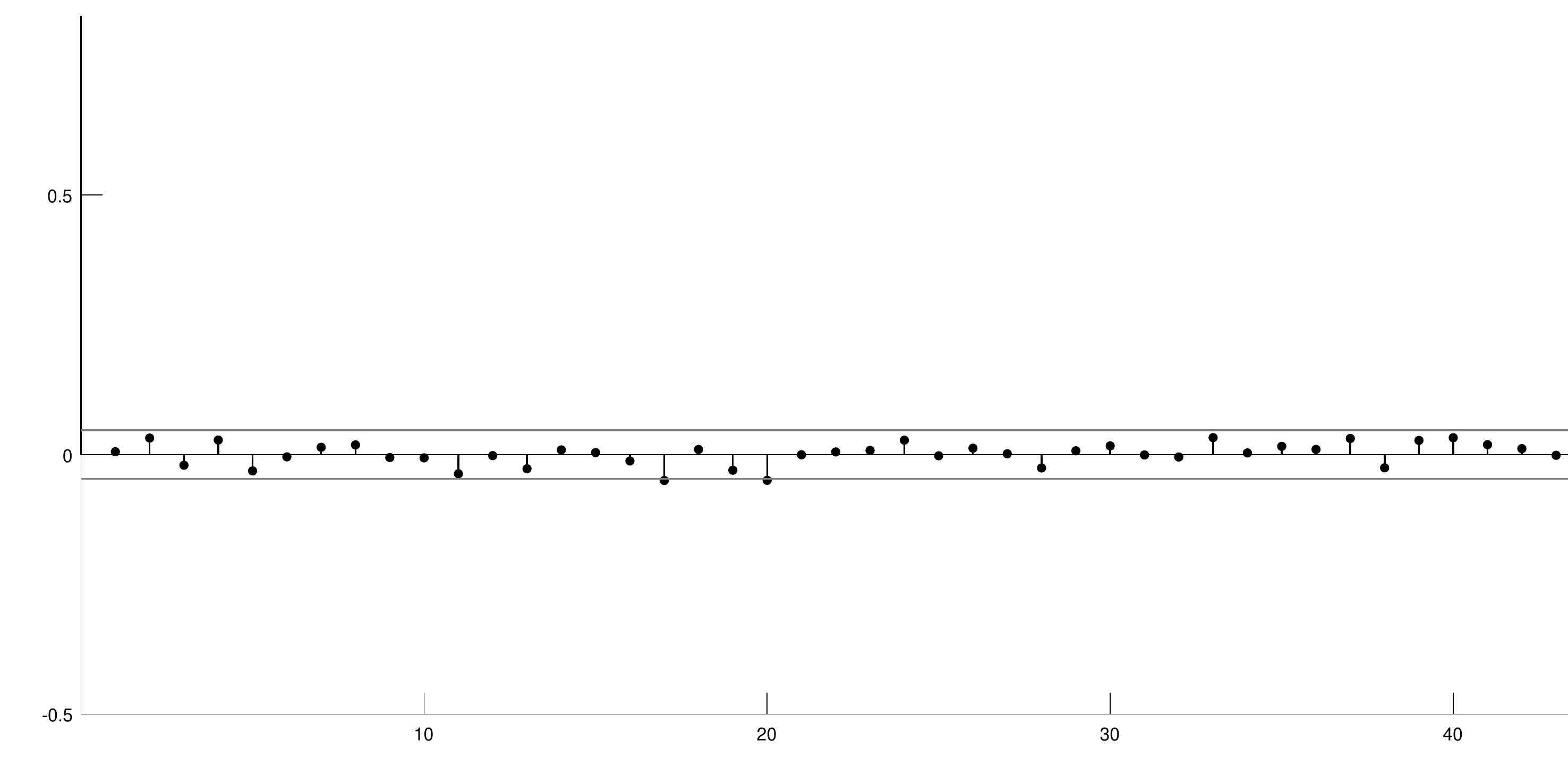}
		\subcaption{Autocorrelation of residuals - VIC}
		\label{fig:vic_r_bi_auto_d}
	\end{subfigure}	
	\caption{Residual analysis of the bivariate shot noise Cox process fitting}
	\label{fig:gof_bivariate}
\end{figure}

Similar to the case of univariate fitting, the goodness-of-fit of the bivariate Cox model is examined by studying the residuals. Here the residuals are defined as the difference between the observed claim counts and the expected claim counts standardised by the standard deviations for all accident days. The empirical distributions and autocorrelations of residuals are presented in Figure \ref{fig:gof_bivariate}. 

\begin{figure}[!ht]
	\centering	
	\begin{subfigure}{0.4\textwidth}
		\centering
		\includegraphics[width=\textwidth]{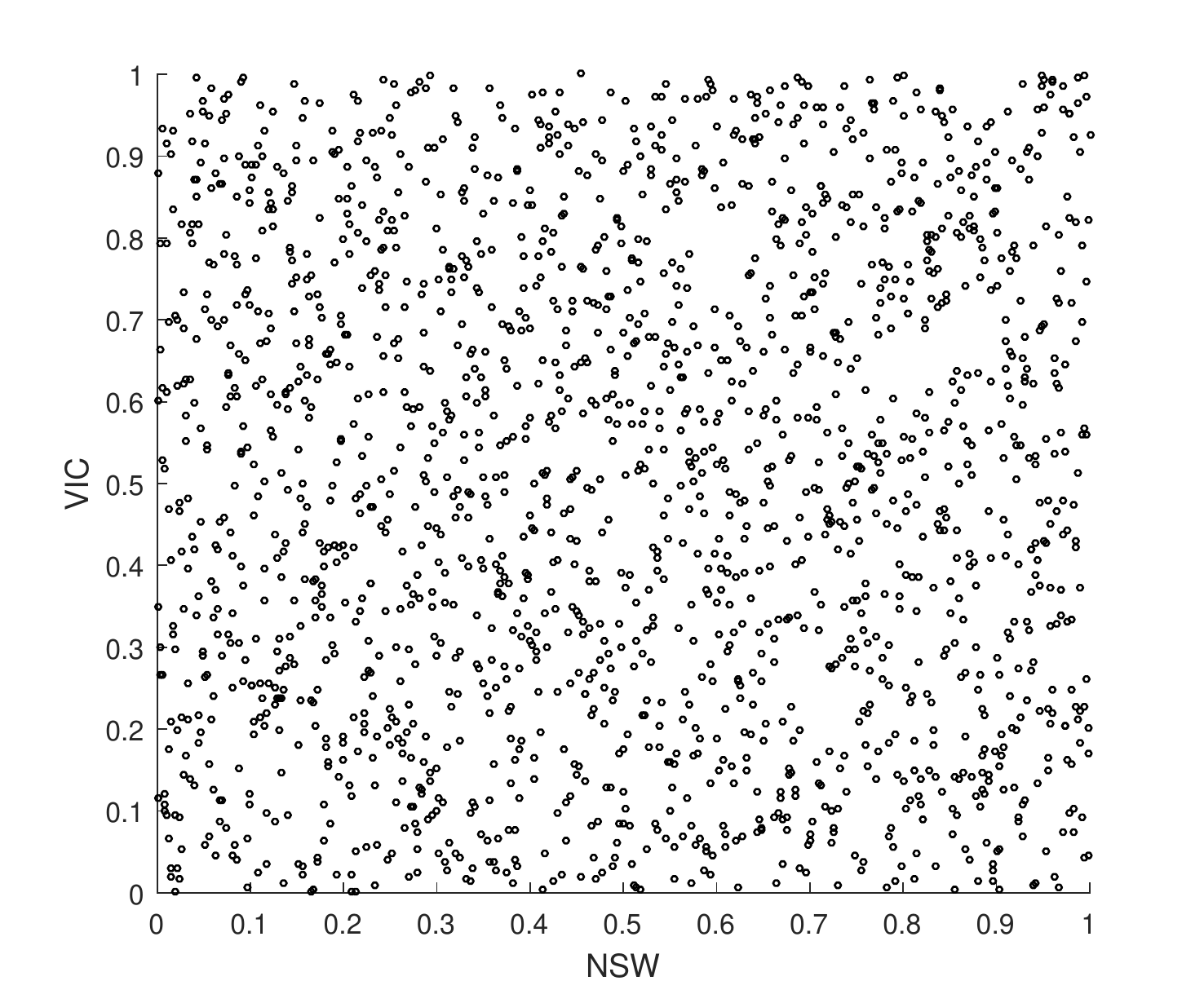}
		\subcaption{Univariate fitting}
		\label{fig:cop_his_mh}
	\end{subfigure}	
	\begin{subfigure}{0.4\textwidth}
		\centering
		\includegraphics[width=\textwidth]{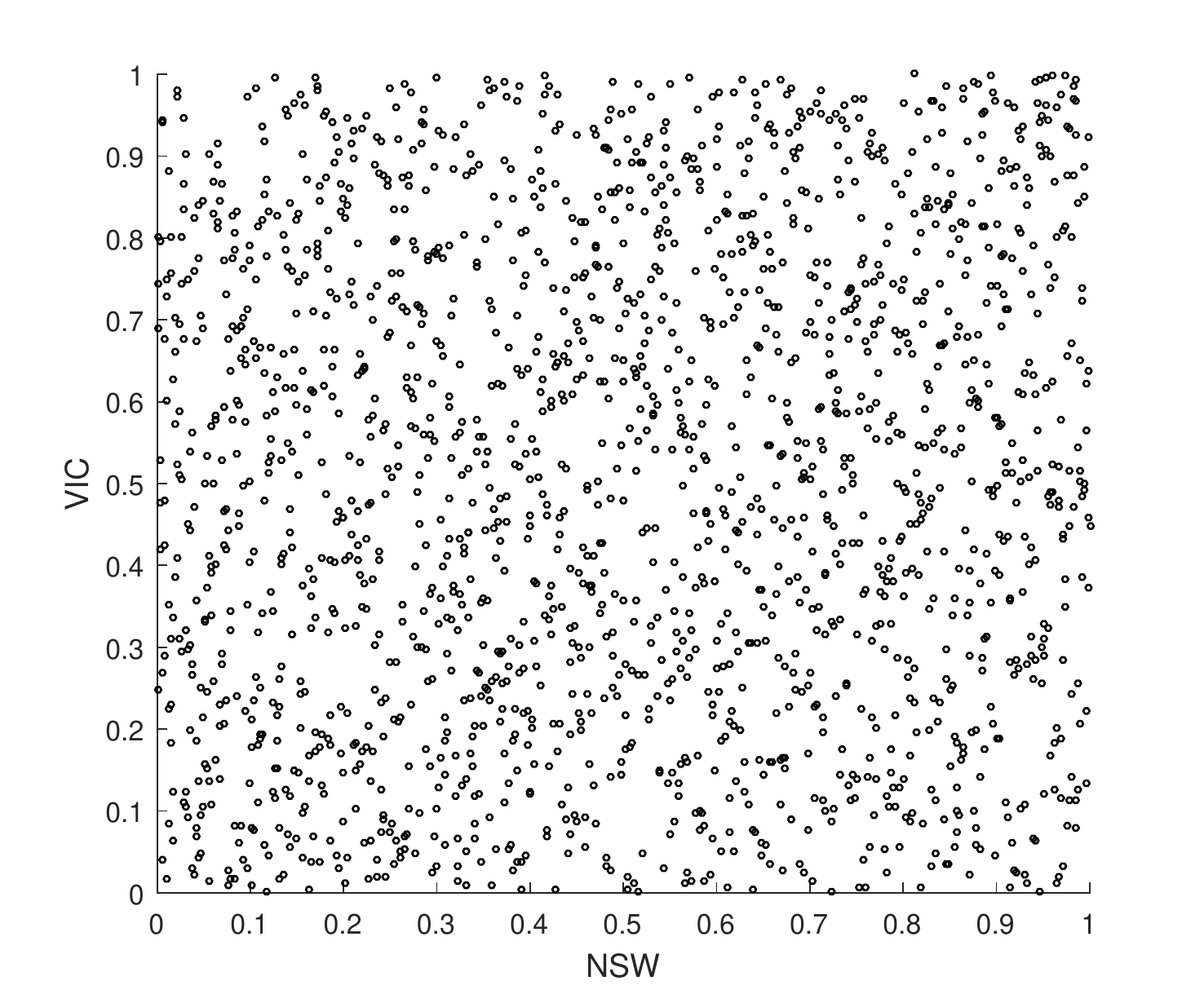}
		\subcaption{Bivariate fitting}
	\end{subfigure}
	\caption{Empirical copulas of \emph{residuals} from fitting the claim arrival components to the NSW and VIC. \emph{Absence of any noticeable dependence structure suggests a good fit.}}
	\label{fig:residual_mh}
\end{figure}

Figure \ref{fig:residual_mh} displays the empirical residual plots of claims of the states, NSW and VIC, from the bivariate fitting. There is no visually significant dependency structure, which suggests good estimation of the underlying intensities and that the shot noise Cox assumptions are appropriate. In particular, Figure \ref{fig:cop_his_mh} shows that the residuals of univariate fitting are rather independent. This suggests that assuming no dependency across the Poisson processes, \emph{given} the intensities, is reasonable. Therefore, despite the fact that the univariate fitting ignores the dependency structure, it filters out the marginal intensities and produces i.i.d. errors. Furthermore, the bivariate filtering decomposes each marginal intensity into a sum of unique and common shot noise. 

\begin{figure}[!ht]
	\centering		
	\begin{subfigure}{0.4\textwidth}
		\centering
		\includegraphics[width=\textwidth]{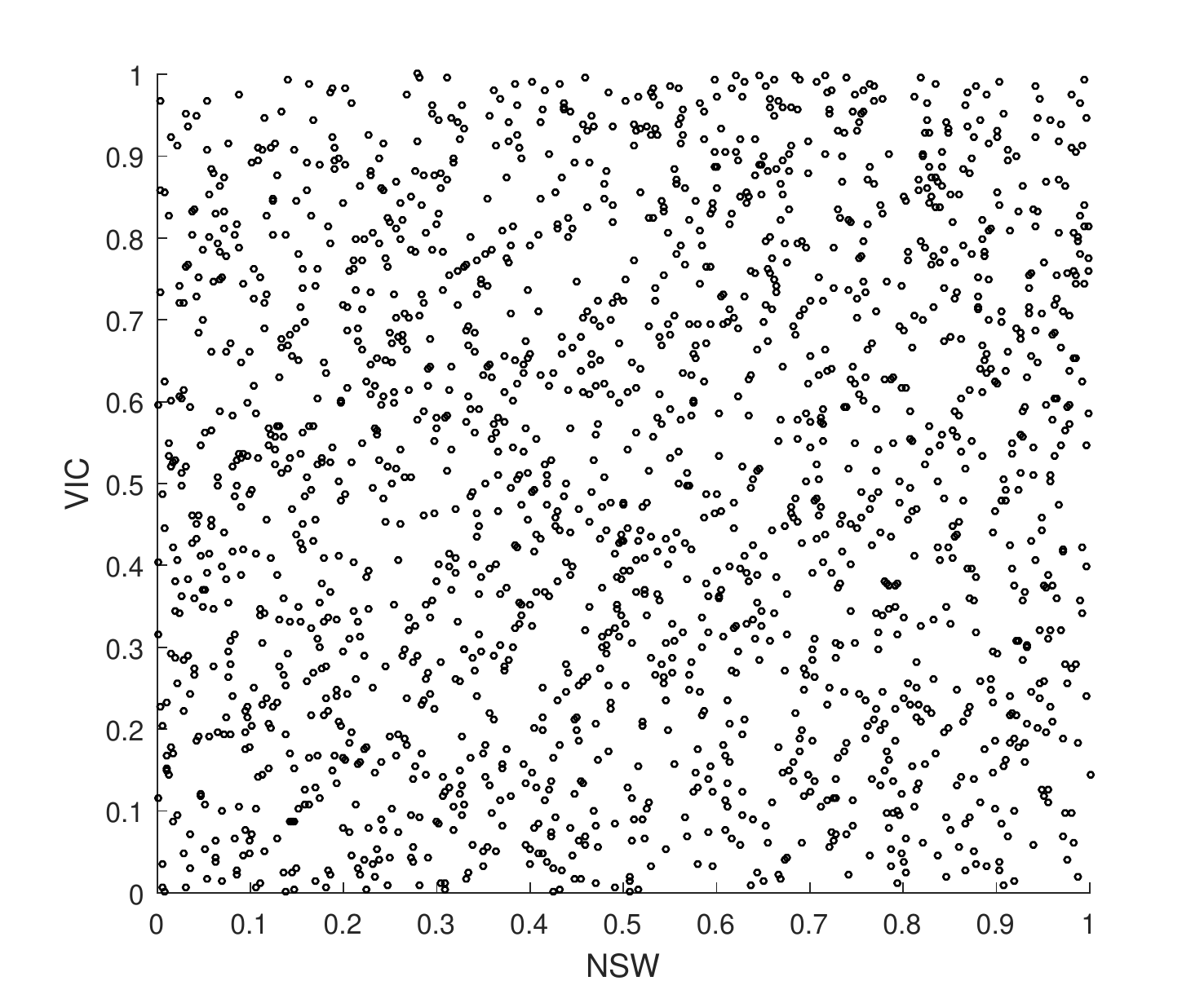}
		\subcaption{Unique components}
		\label{fig:gof_unique}
	\end{subfigure}		
	\begin{subfigure}{0.4\textwidth}
		\centering
		\includegraphics[width=\textwidth]{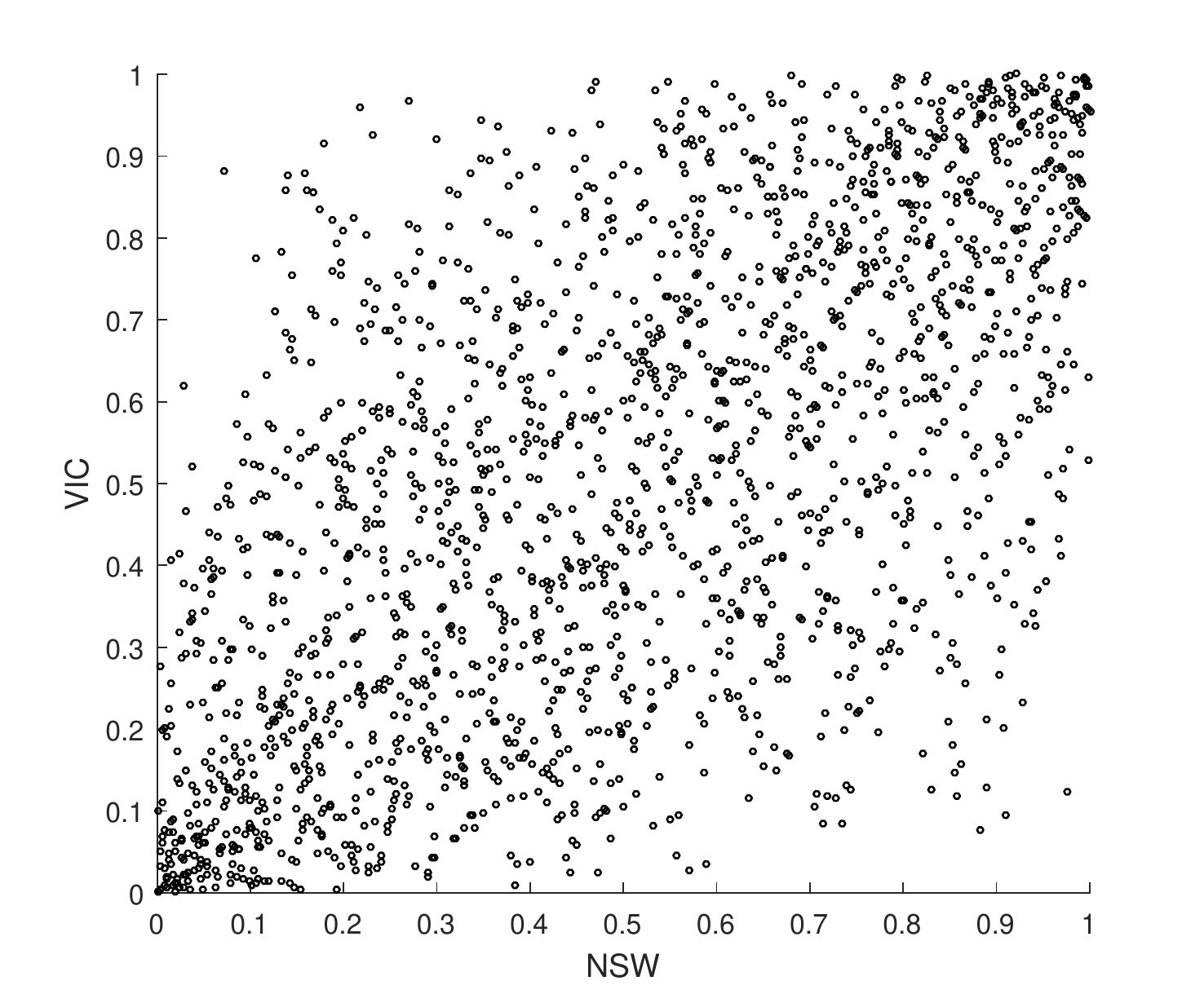}
		\subcaption{Common components}
		\label{fig:gof_common}
	\end{subfigure}

	\caption{Empirical copula of \rev{claim counts (integrated shot noise intensities over each accident day)} of the NSW and VIC states - bivariate fitting. This illustrates the dependence that is present in the data, as filtered and interpreted by our model.}
	\label{fig:copula_intensity_b}
\end{figure}

Figure \ref{fig:copula_intensity_b} displays the empirical copula of the integrated stochastic intensities (over accident days, free of risk exposure) of both states. It is aimed at presenting the dependency between the integrated intensities (that is, the frequencies of daily claims), and therefore the results are filtered intensity processes (which are not observable from data). \rev{This is because realised intensities are not directly observable (only claim counts are), so they must be `inferred' from the claim counts through the lens of a specific model through filtering.} Furthermore, a data point on Figure \ref{fig:copula_intensity_b} refers to the number of daily integrated intensities that fall into each cell, while the integrated intensities are standardised to [0,1]. In particular, Figures \ref{fig:gof_unique} and \ref{fig:gof_common} display the copulas of integrated intensities from the unique jumps and common jumps respectively. This \rev{illustrates} that the L\'evy copula structure further decomposes the marginal intensities into two components, where the unique components are independent and there is a significant positive dependency structure between the common components. 

\begin{remark}
	\rev{As discussed earlier in the paper, it is empirically difficult to judge whether a Cox model is a good candidate directly from the data. However, the model developed in this paper has properties displayed by the data (non-stationary exposure, autocorrelation, \ldots), and it seems that, once fitted, it can explain most of those convincingly. \revv{Specifically, while the exposure adjustment removed much of the (very significant) autocorrelations within both processes (compare the pairs b--d and c--e in both Figures \ref{fig:nsw_count} and \ref{fig:vic_count}), the residual autocorrelations after de-trending are still significantly different from 0 at reasonable confidence level. Furthermore, when comparing Figures \ref{fig:nsw_count}e and \ref{fig:vic_count}e with their counterparts after application of the model in Figures \ref{fig:nsw_r_bi_auto_d} and \ref{fig:vic_r_bi_auto_d}, it appears that those were almost entirely captured by the model.} While we cannot guarantee that our model is the best, we are convinced it is of reasonable quality for this data set.}
\end{remark}

\subsection{Prediction results and discussion}

Given the filtered multivariate intensity, one can investigate further how the unique and common shots contribute to the integrated stochastic intensities. The results in Table \ref{tab:decomposition} indicate substantial weights from common shocks in both marginal intensities. For VIC, the common events contribution to 49.30\% of the stochastic intensities of the claim arrival process. Here 49.30\% does not refer to the number of events; instead, it refers to the combination effects of frequency, severity and decay of the shot events on the stochastic intensity of claim arrivals in the state of VIC.

\begin{table}[htb]
	\centering
	\begin{tabular}{p{7cm} p{3cm} p{3cm}}
		\hline
		& NSW               & VIC                \\ \hline
		Contribution from unique shot noise    & 67.54\%             & 50.70\%             \\
		Contribution from common shot noise    & 32.46\%             & 49.30\%             \\ \hline
	\end{tabular}
	\caption{Decomposition of the integrated intensities}
	\label{tab:decomposition}
\end{table}

\begin{figure}[!ht]
	\centering	
	\includegraphics[width=0.75\textwidth]{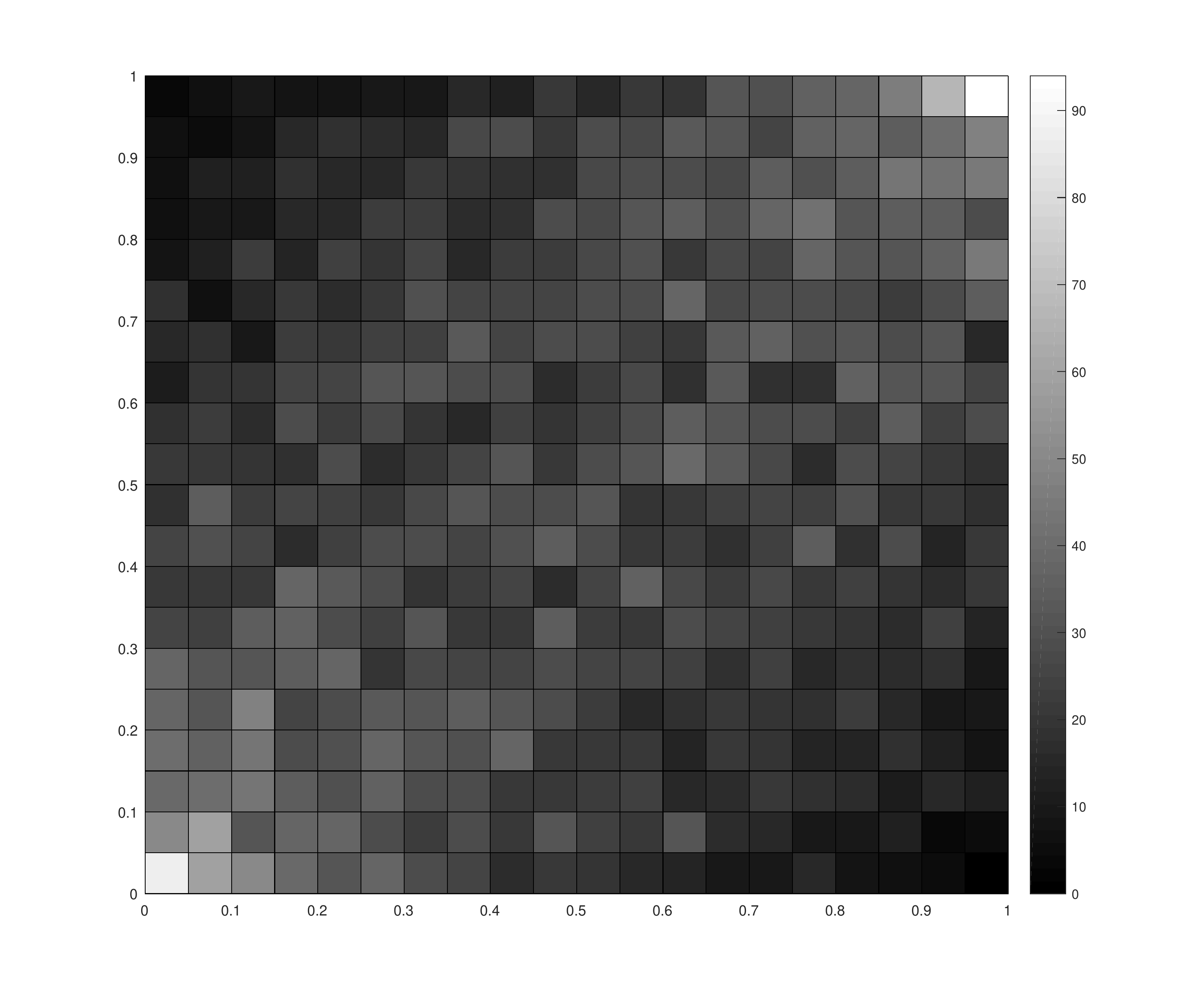}	
	\caption{Heat maps of the histograms of the empirical copulas between the simulated claims counts}
	\label{fig:prediction}
\end{figure}

We predicted the distributions of total future claim counts over the next one year (assuming constant risk exposures) through 100,000 simulations. Note that the initial value of the shot noise trajectory is chosen as the filtered shot noise intensity at the end of the observation period, hence the simulations of future scenarios are implicitly based on the latest status of the shot noise development. Figure \ref{fig:prediction} visualises the bivariate histogram of the copulas between claim counts in the two cases. Here the \rev{colour intensity} in each cell refers to the number of simulations where the bivariate claim count fall into a specific joint quantile range. It shows that there is significant evidence of dependency, which is consistent with the results in Table \ref{tab:decomposition}. This is further confirmed with the numerical dependency measures in Table \ref{tab:dependency_measure}, where all the measures are material and statistically significant (with all $p$-values being almost 0). Note that the values in Table \ref{tab:dependency_measure} should not be compared directly to dependency measures of real data. This is due to the non-constant risk exposures and also serial dependency of claim counts over time, which means the i.i.d. assumptions of claim counts in constructing the dependency measures is not valid.

The dependency measures in Table \ref{tab:dependency_measure} refer to the dependency structure of the next year, based on a given set of controlled factors. Our methodology provides a statistical sound way of making assumptions about the dependency structures. The implied future dependency measures can then be used in more traditional (and straightforward) methodologies, for example, when a correlation matrix is required to aggregate individual portfolios to a company level. This improves the existing practice of multivariate reserving where dependency structures are usually based on expert knowledge and industry statistics. \rev{ In particular, one common approach to estimate risk margin at a company level is by aggregating individual risk margins of various portfolios with the help of correlations. But correlations typically cannot be inferred from data (as 10 years of data would yield only 10 observations, which is insufficient to estimate yearly correlations), and are hence generally chosen judgmentally \citep*[see also][for further discussion of this]{AvTaWo16}. Our approach provides a promising first step towards estimating dependency measures through a more rigorous and objective approach. Indeed, only a few years of data are required to fit a model that can subsequently yield implied dependence measures for any time horizon.}

\begin{table}[htb]
	\centering
	\begin{tabular}{ p{4cm} p{3cm}l}
		\hline
		Pearson's correlation & Kendall's tau & Spearman's rho \\
		0.3633                & 0.2355       & 0.3469         \\ \hline		
	\end{tabular}
	\caption{Dependency measurement across the bivariate prediction of claims counts}
	\label{tab:dependency_measure}
\end{table}

\begin{remark}
	\rev{We have also attempted to provide comparison between our model and the Chain Ladder model, however we realise that such a comparison can be challenging and may provide misleading results.} 
	There are fundamental differences in the theoretical properties and how the models are implemented between the Chain Ladder model (and most of the existing reserving models) and our approach. These facts make it very difficult to conduct a meaningful comparison. \revv{That being said, the multivariate Cox model can be complementary to the Chain Ladder model. While the Chain Ladder model is commonly used to estimate the risk margins of individual LoBs, our multivariate Cox model can provide insights into the dependence structures at play, and hence help modellers make informed and educated choice for the correlations that would be used in traditional modelling approaches to aggregate results.}

\end{remark}

\section{Conclusion} \label{S_conclusion}

\revv{In this paper, we have developed a \emph{multivariate} Cox shot noise process approach to model (claim) counts. Common shocks were introduced on the intensities of claim arrivals across multiple \revv{counts processes} with the help of L\'evy copulas. We also developed a filtering algorithm to estimate the underlying joint intensity.}

\rev{The independent increment property of a L\'evy process means the dependency structure of a multivariate compound Poisson process can only be introduced via a common shock model. However, there is a bijective relationship between a L\'evy copula model and a common shock model \citep*[see Sklar's theorem for L\'evy copulas in][for example]{Tank03}. Hence, our proposed approach is actually not different from a common shock approach, but it is significantly more tractable.} \revv{It also has the advantage of introducing dependence \emph{within} each process, a common feature observed in actuarial data sets.}

The model construction and calibration procedures are illustrated with a real bivariate data set. In particular, we account for time covariates (including weekly patterns, annual patterns, and trends). After allowing for these covariates, the results show that the dependency between the claim arrival processes of the Motor LoB in both NSW and VIC states is still significant. This is expected\rev{, due to random phenomena that may affect both states at the same time (e.g., weather events)}. \revv{The Cox model presented in this paper proved successful in capturing the residual correlation structures that were left in the data after detrending.}

\revv{Our work focused on the particular issue of dependency modelling in claim frequency. Such a framework may help in understanding the dependency structure between the claim processes of multiple LoBs, and in particular, its impact on the quantiles of the aggregated loss.} For example, one implementation of the multivariate Cox process is to derive \revv{an} implied \rev{dependence} structure (see \rev{e.g.} Figure \ref{fig:prediction}), which can be utilised in aggregating individual reserve estimates to a company level. 

\revv{The model implementation procedures developed in this paper can be easily extended from our bivariate illustration. This will require an appropriate choice of a higher dimension L\'evy copula. Should the dependency structure across multiple counts processes be non-exhangeable, appropriate L\'evy copulas can be constructed \citep*[see, for example][]{AvTaWoYa16}.} 


\section*{Acknowledgements}

\rev{Authors are very grateful for comments from two referees, which led to significant improvements of the paper.} Earlier versions of this paper were presented at the $20^\text{th}$ International Congress on Insurance: Mathematics and Economics (Atlanta, U.S.A.), 
at the $3^\text{rd}$ European Actuarial Journal Conference (Lyon, France), 
at the $22^\text{nd}$ International Congress on Insurance: Mathematics and Economics (Sydney, Australia), 
and at the Australian National University Research School of Finance, Actuarial Studies and Statistics 2018 Summer Camp. 
The authors are grateful for constructive comments received from colleagues who attended those events. 

This research was supported under Australian Research Council's Linkage (LP130100723, with funding partners Allianz Australia Insurance Ltd, Insurance Australia Group Ltd, and Suncorp Metway Ltd) and Discovery (DP200101859) Projects funding schemes. Furthermore, Avanzi acknowledges support from a grant of the Natural Science and Engineering Research Council of Canada (project number RGPIN-2015-04975). Finally, Yang acknowledges financial support from an Australian Postgraduate Award and supplementary scholarships provided by the UNSW Business School. The views expressed herein are those of the authors and are not necessarily those of the supporting organisations. 

\bibliographystyle{elsarticle-harv}
\bibliography{libraries}

\appendix

\section{Proof for Example \ref{exe:bivariate}} \label{append:proof}
\begin{proof}
	Consider the process of the common shots, we have 
	\begin{equation}
	\begin{aligned}
	\coml{1}(t)&=\coml{1}(0)e^{-t\kappa_1}+\sum_{i=1}^{\comj{12}(t)}\comx{i,1:12}e^{-(t-\comt{i,12})\kappa_1},\\
	\coml{2}(t)&=\coml{2}(0)e^{-t\kappa_2}+\sum_{i=1}^{\comj{12}(t)}\comx{i,2:12}e^{-(t-\comt{i,12})\kappa_2}.
	\end{aligned}
	\end{equation}
	
	We characterise the trajectory of the bivariate shot noise process by using a random vector, $\comtheta{12}{t}$, which is the collection of the number of jumps, the location and joint severities of each shot up to time $t$, that is
	
	\begin{equation}
	\begin{aligned}
	\comtheta{12}{t}=\{\comj{12}(t), \comt{1,12}, \ldots, \comt{\comj{12}(t),12}, \comx{1,12}, \ldots, \comx{\comj{12}(t),12} \}.
	\end{aligned}
	\end{equation}
	
	Based on the Conditional Covariance Formula \citep*[see, for example][Chapter 7]{Ros14}, we have 
	
	\begin{equation}
	\begin{aligned}
	\cov{\coml{1}(t),\coml{2}(t)}=\E{\cov{\coml{1}(t),\coml{2}(t)\cond\comj{12}(t)}}+\cov{\E{\coml{1}(t)\cond\comj{12}{t}},\E{\coml{2}(t)\cond\comj{12}(t)}}.
	\end{aligned}
	\end{equation}
	
	Furthermore, conditional on $\comj{12}{(t)}$ (which is a Poisson random variable itself with intensity $\comr{12}$), the locations of shots $\left(\text{that is, } \{\comt{i,12}\cond\comj{12}(t)\}_{i=1,\ldots,\comj{12}(t)}\right)$ are independent uniform random variables.
	
	Firstly, we start from deriving $\E{\cov{\coml{1}(t),\coml{2}(t)\cond\comj{12}(t)}}$. We have:
	
	\begin{equation}
	\begin{aligned}
	&\E{\cov{\coml{1}(t),\coml{2}(t)\cond\comj{12}(t)}}\\
	=&\E{\cov{\coml{1}(0)e^{-t\kappa_1}+\sum_{i=1}^{\comj{12}(t)}\comx{i,1:12}e^{-(t-\comt{i,12})\kappa_1},\coml{2}(0)e^{-t\kappa_2}+\sum_{i=1}^{\comj{12}(t)}\comx{i,2:12}e^{-(t-\comt{i,12})\kappa_2}\cond\comj{12}(t)}}\\
	=&\E{e^{-t(\kappa_1+\kappa_2)}\cov{\coml{1}(0),\coml{2}(0)}+\comj{12}(t)\cov{\comx{1:12}e^{-(t-\comt{12})\kappa_1},\comx{2:12}e^{-(t-\comt{12})\kappa_1}\cond \comj{12}(t)}}\\
	=&\E{e^{-t(\kappa_1+\kappa_2)}\cov{\coml{1}(0),\coml{2}(0)}+\comj{12}(t)e^{-t(\kappa_1+\kappa_2)}\cov{\comx{1:12}e^{\comt{12}\kappa_1},\comx{1:12}e^{\comt{12}\kappa_2}\cond \comj{12}(t)}}\\
	\end{aligned}
	\end{equation}
	and
	\begin{equation}
	\begin{aligned}
	&\cov{\comx{1:12}e^{\comt{12}\kappa_1},\comx{1:12}e^{\comt{12}\kappa_2}\cond \comj{12}(t)}\\
	=&\E{\comx{1:12}e^{\comt{12}\kappa_1}\comx{1:12}e^{\comt{12}\kappa_2}|\comj{12}(t)}-\E{\comx{1:12}e^{\comt{12}\kappa_1}\cond\comj{12}(t)}\E{\comx{1:12}e^{\comt{12}\kappa_2}\cond\comj{12}(t)}\\
	=&\E{\comx{1:12}\comx{2:12}}\E{e^{\comt{12}(\kappa_1+\kappa_2)}\cond \comj{12}(12)}-\E{\comx{1:12}}\E{\comx{2:12}}\E{e^{\comt{12}\kappa_1}\cond \comj{12}(t)}\E{e^{\comt{12}\kappa_2}\cond \comj{12}(t)}.
	\end{aligned}
	\end{equation}
	
	The location of the shot given the total number of shot is uniformly distributed over $[0,t]$, therefore one can obtain the first two moments by:
	
	\begin{equation}
	\begin{aligned}
	\E{e^{\comt{12}a}\cond \comj{12}(t)}&=\int_{0}^{t}e^{ax}\frac{1}{t}\;\mathrm{d}x=\frac{e^{at}-1}{at},\\
	\end{aligned}
	\end{equation}
	
	hence
	\begin{equation}
	\begin{aligned}
	&\cov{\comx{1:12}e^{\comt{12}\kappa_1},\comx{1:12}e^{\comt{12}\kappa_2}\cond \comj{12}(t)}\\
	=&\E{\comx{1:12}\comx{2:12}}\frac{e^{(\kappa_1+\kappa_2)t}-1}{(\kappa_1+\kappa_2)t}-\E{\comx{1:12}}\E{\comx{2:12}}\frac{e^{\kappa_1t}-1}{\kappa_1t}\frac{e^{\kappa_2t}-1}{\kappa_2t}.
	\end{aligned}
	\end{equation}
	
	Therefore,
	
	\begin{equation}
	\begin{aligned}
	&\E{\cov{\coml{1}(t),\coml{2}(t)\cond\comj{12}{t}}}\\
	=&e^{-t(\kappa_1+\kappa_2)}\cov{\coml{1}(0),\coml{2}(0)}\\
	+&\comr{12}te^{-t(\kappa_1+\kappa_2)}\left(\E{\comx{1:12}\comx{2:12}}\frac{e^{(\kappa_1+\kappa_2)t}-1}{(\kappa_1+\kappa_2)t}-\E{\comx{1:12}}\E{\comx{2:12}}\frac{e^{\kappa_1t}-1}{\kappa_1t}\frac{e^{\kappa_2t}-1}{\kappa_2t}\right).
	\end{aligned}
	\end{equation}
	
	Secondly, we derive $\cov{\E{\coml{1}(t)\cond\comj{12}(t)},\E{\coml{2}(t)\cond\comj{12}(t)}}$. We start from deriving the conditional expectation terms:
	
	\begin{equation}
	\begin{aligned}
	\E{\coml{1}(t)\cond\comj{12}(t)}&=\E{\coml{1}(0)e^{-t\kappa_1}+\sum_{i=1}^{\comj{12}(t)}\comx{1:12}e^{-(t-\comt{12})\kappa_1}\cond\comj{12}(t)}\\
	&=e^{-t\kappa_1}\left(\E{\coml{1}(0)}+\comj{12}(t)\E{\comx{1:12}}\E{e^{\comt{12}\kappa_1}}\right)\\
	&=e^{-t\kappa_1}\left(\E{\coml{1}(0)}+\comj{12}(t)\E{\comx{1:12}}\frac{e^{\kappa_1t}-1}{\kappa_1t}\right).
	\end{aligned}
	\end{equation}
	
	\begin{equation}
	\begin{aligned}
	\E{\coml{2}(t)\cond\comj{12}{t}}=e^{-t\kappa_2}\left(\E{\coml{2}(0)}+\comj{12}(t)\E{\comx{2:12}}\frac{e^{\kappa_2t}-1}{\kappa_2t}\right)
	\end{aligned}
	\end{equation}
	
	Therefore we arrive at
	
	\begin{equation}
	\begin{aligned}
	&\cov{\E{\coml{1}(t)\cond\comj{12}{t}},\E{\coml{2}(t)\cond\comj{12}{t}}}\\
	=&e^{-t(\kappa_1+\kappa_2)}\frac{e^{\kappa_1t}-1}{\kappa_1t}\frac{e^{\kappa_2t}-1}{\kappa_2t}\E{\comx{1:12}}\E{\comx{2:12}}\var{\comj{12}(t)}\\
	=&e^{-t(\kappa_1+\kappa_2)}\frac{e^{\kappa_1t}-1}{\kappa_1t}\frac{e^{\kappa_2t}-1}{\kappa_2t}\comr{12}t\E{\comx{1:12}}\E{\comx{2:12}}\\	
	\end{aligned}
	\end{equation}
	
	and 
	
	\begin{equation}
	\begin{aligned}
	&\cov{\coml{1:12}(t),\coml{2:12}(t)}\\
	=&e^{-t(\kappa_1+\kappa_2)}\cov{\coml{1}(0),\coml{2}(0)}\\
	+&\comr{12}te^{-t(\kappa_1+\kappa_2)}\left(\E{\comx{1:12}\comx{2:12}}\frac{e^{(\kappa_1+\kappa_2)t}-1}{(\kappa_1+\kappa_2)t}-\E{\comx{1:12}}\E{\comx{2:12}}\frac{e^{\kappa_1t}-1}{\kappa_1t}\frac{e^{\kappa_2t}-1}{\kappa_2t}\right)\\
	+&e^{-t(\kappa_1+\kappa_2)}\frac{e^{\kappa_1t}-1}{\kappa_1t}\frac{e^{\kappa_2t}-1}{\kappa_2t}\comr{12}t\E{\comx{1:12}}\E{\comx{2:12}}\\
	=& e^{-t(k_1+k_2)}\cov{\coml{1}(0),\coml{2}(0)} + \comr{12}te^{-t(\kappa_1+\kappa_2)}\left(\E{\comx{1:12}\comx{2:12}}\frac{e^{(\kappa_1+\kappa_2)t}-1}{(\kappa_1+\kappa_2)t} \right).
	\end{aligned}
	\end{equation}
	
	Given that this is a stationary bivariate process, we have:
	 
	\begin{equation}
	\cov{\coml{1:12}(t),\coml{2:12}(t)}=\cov{\coml{1:12}(0),\coml{2:12}(0)},
	\end{equation} 
	
	hence
	
	\begin{equation}
	\begin{aligned}
	&\left(1-e^{-t(\kappa_1+\kappa_2)}\right)\cov{\coml{1:12}(t),\coml{2:12}(t)}\\
	=&\comr{12}te^{-t(\kappa_1+\kappa_2)}\E{\comx{1:12}\comx{2:12}}\frac{e^{(\kappa_1+\kappa_2)t}-1}{(\kappa_1+\kappa_2)t}.
	\end{aligned}
	\end{equation}
	
	Therefore 
	
	\begin{equation}
	\begin{aligned}
	\cov{\coml{1:12}(t),\coml{2:12}(t)}=\frac{\comr{12}\E{\comx{1:12}\comx{2:12}}}{\kappa_1+\kappa_2}.
	\end{aligned}
	\end{equation}
	
	Since the two unique shot processes are independent with the bivariate common shot process, therefore
	
	\begin{equation}
	\begin{aligned}
	\cov{\tilde{\lambda}_1(t),\tilde{\lambda}_2(t)}=\cov{\coml{1:12}(t)+\unil{1}(t),\coml{2:12}(t)+\unil{2}(t)}=\cov{\coml{1:12}(t),\coml{2:12}(t)},
	\end{aligned}
	\end{equation}
	which completes the proof.
	\end{proof}

\end{document}